\documentclass[conference]{IEEEtran}
\usepackage{tabularx}
\usepackage{cite}
\usepackage{amsmath,amssymb,amsfonts}
\usepackage{algorithmic}
\usepackage{graphicx}
\usepackage{textcomp}
\usepackage{bmpsize}
\usepackage{xcolor}
\usepackage{lipsum}
\usepackage[colorlinks=true,urlcolor=black]{hyperref}
\def\BibTeX{{\rm B\kern-.05em{\sc i\kern-.025em b}\kern-.08em
    T\kern-.1667em\lower.7ex\hbox{E}\kern-.125emX}}
\usepackage{url} 
\usepackage{tikz} 
\usepackage{varwidth} 
\usepackage{listings}
\newtheorem{theorem}{Theorem}[section]
\newtheorem{remark}{Remark}
\newtheorem{definition}[theorem]{Definition}
\newtheorem{proposition}[theorem]{Proposition}
\newcommand{\comment}[2]{#2}

\usepackage{dcolumn}
\usepackage{booktabs}
\usetikzlibrary{positioning,arrows,shapes.misc,shapes.gates.logic.US,calc,shapes}
\newcolumntype{M}[1]{D{.}{.}{1.#1}}
\usepackage[normalem]{ulem}
\usepackage{amsmath, amssymb}
\usepackage[boxed]{algorithm2e}
\SetKwProg{Fn}{def}{\string:}{}
\SetKwFunction{FRecurs}{RecursiveProbability}%
\SetKw{KwTo}{in}\SetKwFor{For}{for}{\string:}{}%
\SetKwIF{If}{ElseIf}{Else}{if}{:}{elif}{else:}{}%
\SetKwFor{While}{while}{:}{fintq}%
\SetAlgoNoEnd\SetAlgoNoLine%
\DontPrintSemicolon
\makeatletter
\newcommand{\removelatexerror}{\let\@latex@error\@gobble}
\makeatother
\begin{document}

\title{Cyclic Bayesian Attack Graphs: A Systematic\\ Computational Approach}

\author{\IEEEauthorblockN{Isaac Matthews}
\IEEEauthorblockA{
\textit{Newcastle University}\\
Newcastle upon Tyne, U.K.\\
I.J.Matthews2@ncl.ac.uk}
\and
\IEEEauthorblockN{John Mace}
\IEEEauthorblockA{
\textit{Newcastle University}\\
Newcastle upon Tyne, U.K.}
\and
\IEEEauthorblockN{Sadegh Soudjani}
\IEEEauthorblockA{
\textit{Newcastle University}\\
Newcastle upon Tyne, U.K.}
\and
\IEEEauthorblockN{Aad van Moorsel}
\IEEEauthorblockA{
\textit{Newcastle University}\\
\quad Newcastle upon Tyne, U.K.\quad}
}

\maketitle

\begin{abstract}
Attack graphs are commonly used to analyse the security of medium-sized to large networks. Based on a scan of the network and likelihood information of vulnerabilities, attack graphs can be transformed into Bayesian Attack Graphs (BAGs).  These BAGs are used to evaluate how security controls affect a network and how changes in topology affect security.
A challenge with these automatically generated BAGs is that cycles arise naturally, which make it impossible to use Bayesian network theory to calculate state probabilities.  In this paper we provide a systematic approach to analyse and perform computations over cyclic Bayesian attack graphs. 
Our approach first formally introduces two commonly used versions of Bayesian attack graphs and compares their expressiveness.  We then present an interpretation of Bayesian attack graphs based on combinational logic circuits, which facilitates an intuitively attractive systematic treatment of cycles. We prove properties of the associated logic circuit and present an algorithm that computes state probabilities without altering the attack graphs (e.g., remove an arc to remove a cycle). Moreover, our algorithm deals seamlessly with all cycles without the need to identify their types.
A set of experiments using synthetically created networks demonstrates the scalability of the algorithm on computer networks with hundreds of machines, each with multiple vulnerabilities.  
\end{abstract}


\section{Introduction}




An attack graph is a representation of a system and its vulnerabilities in the form of a directed acyclic graph. It models how a system's vulnerabilities can be leveraged during a single attack to progress through a network. In recent years, several authors have pursued to combine Bayesian statistics with attack graphs to automatically prioritise network vulnerabilities from a probabilistic view point, resulting in Bayesian Attack Graphs \cite{Aguessy2016,Huangfu2017,Munoz-Gonzalez2017,Ramaki2015,sembiring2015network,Dantu2004,Doynikova2017,homer2013aggregating}.  The literature discusses a variety of considerations when generating BAGs \cite{Miehling2015:ODP:2808475.2808482,Xie2010}, including using the Common Vulnerability Scoring System (CVSS) to associate probabilities with vulnerabilities \cite{Frigault2017}. This approach has received uptake in practical security analysis systems, in particular in MulVAL \cite{Ou2005:MLN:1251398.1251406}, which automatically generate BAGs from network scans and CVSS information.  

Once a BAG has been generated it can be analysed in various ways. One approach is an exclusively static analysis, in order to identify the weakest areas of a network as well as identify quantitatively the risk that a certain asset will be compromised in the case of an attack. Further extensions to this kind of analysis include the introduction of attack profiles to modify the probabilities; for example this could be done using the attack complexity metric in the CVSS base vector to determine the ease of an attack. One approach to this is detailed by Cheng et al.~\cite{Wang2017networkbook} along with a method to include dependency relationships between vulnerabilities. Another application of the BAG is as a dynamic risk assessment tool where an administrator can model new security controls and their effects on a network, as well as dynamically analyse a deployed network's most likely attack paths, that can be updated dependent on information from an intrusion detection system \cite{Poolsappasit2012}.



The majority of the techniques used to calculate probabilities for BAGs require that they do not contain `cycles' but are allowed to have `loops' \cite{Frigault2017,liu2005network,Miehling2015:ODP:2808475.2808482}. Such acyclic BAGs follow the \emph{monotonicity principle}, that an attacker will never return to a previous state. However, networks that arise in practice when using tools such as MulVAL routinely contain cycles. These cycles arise naturally, as we will illustrate in Section \ref{sec:running_example} for a canonical example.


To deal with cycles in BAGs, the existing literature suggest to remove edges from the graph to prevent the attacker backtracking \cite{Frigault2017,Saha2008,Doynikova2017}.  There are a number of practical drawbacks to this approach, especially when carried out outside the analysis algorithm, thus altering the model.  For instance, removing an edge can make reasoning about the graph for a cyber security professional confusing, if the graph is examined by hand to identify specific routes and there are edges missing for the calculations.  In addition, avoiding cycles that occur when generating the BAG could be impractical and take a ``substantial amount of time" \cite{Swiler2001} to ensure an edge is not cycle-causing whenever a new edge is being added to the graph.

In this paper we propose a systematic computational approach for analysing cyclic BAGs, combining formalising models (attack graphs, BAGs and variants) and their properties throughout. We integrate the resolution of cycles in the algorithm for computing state probabilities of the BAG.  The benefit of our approach is twofold.  First, we establish a more formal base for the discussion of attack graphs, (cyclic) BAGs and solution techniques for BAGs, something that has not been strongly developed in the literature thus far. Secondly, we provide a single unified solution algorithm for BAGs that resolves the problem of cycles for any cyclic BAG, without altering the attack graph.

Our most significant contribution is a single unified solution to the static analysis of any BAG that does not modify the graph in any way while being able to run on graphs containing cycles. This can be used to properly analyse security threats to a network and correctly prioritise remediation steps.
Moreover, when applied to acyclic BAGs, our solution provides exactly the same results identical with the outcome of approaches in the literature that deal only with the acyclic case, thus is a generalisation of these approaches.

More specifically, we first formalize two common types of attack graphs, and compare their expressiveness. The AND/OR style of attack graphs is the most powerful and is used subsequently throughout the paper. We then introduce a novel interpretation and formalization of attack graphs using combinational logic circuits that helps us reason about cycles more effectively.  Combinational circuits allows one to capture intuitively the notion of subsequent visits to the same state, which we use to reason about cycles.
We show that the types of cycles studied in the literature correspond to different ways in which probabilities change as a function of the visit count of a state.

Based on the formalization and the associated intuition, we derive an algorithm that handles cycles in BAGs regardless of their type in a natural manner.  Due to the reasons mentioned above, the algorithm does not explicitly identify the edges in the attack graph that should be removed.  Instead, it integrates the identification of cycles with the solution algorithm and then terminates the recursion.  The algorithm is suitable for solving state probabilities in the BAG, and gives results for acyclic BAGs that are identical with the output of traditional solvers for acyclic BAGs.

To study the scalability of our algorithm, we generate synthetic BAGs of various size, in a manner similar to \cite{Munoz-Gonzalez2017}. We conclude that the algorithm can be used with BAGs of size 15000 nodes with reasonable computation time, implying it can scale to be used with networks of at least 750 hosts each with 5 vulnerabilities.

The rest of this paper is organised as follows. Section~\ref{sec:motivation} introduces the network architecture that will be used as a running example throughout the paper. It also provides the motivation for our contributions and discusses two common formalisms for the construction of BAGs, then relates and shows the equivalence between the two formalisms.
Section~\ref{sec:access} introduces Bayesian networks and relates the process of Variable Elimination for Bayesian networks to the calculation of probabilities in acyclic BAGs. Section~\ref{sec:combinational} shows how an attack graph can be interpreted as a deterministic combinational circuit with probabilistic inputs. Section~\ref{sec:calculation} presents our unified solution for dealing with both cyclic and acyclic BAGs, and section~\ref{sec:experiments} details the experiments run for our solution on both common and simulated examples. Finally, related works and our conclusions are presented in Sections~\ref{sec:related}~and~\ref{sec:conclusion}, respectively.

\section{Motivation and Problem Formulation}
\label{sec:motivation}

\subsection{Running Example}
\label{sec:running_example}

\begin{figure}
	\centering
	\includegraphics[width=\linewidth]{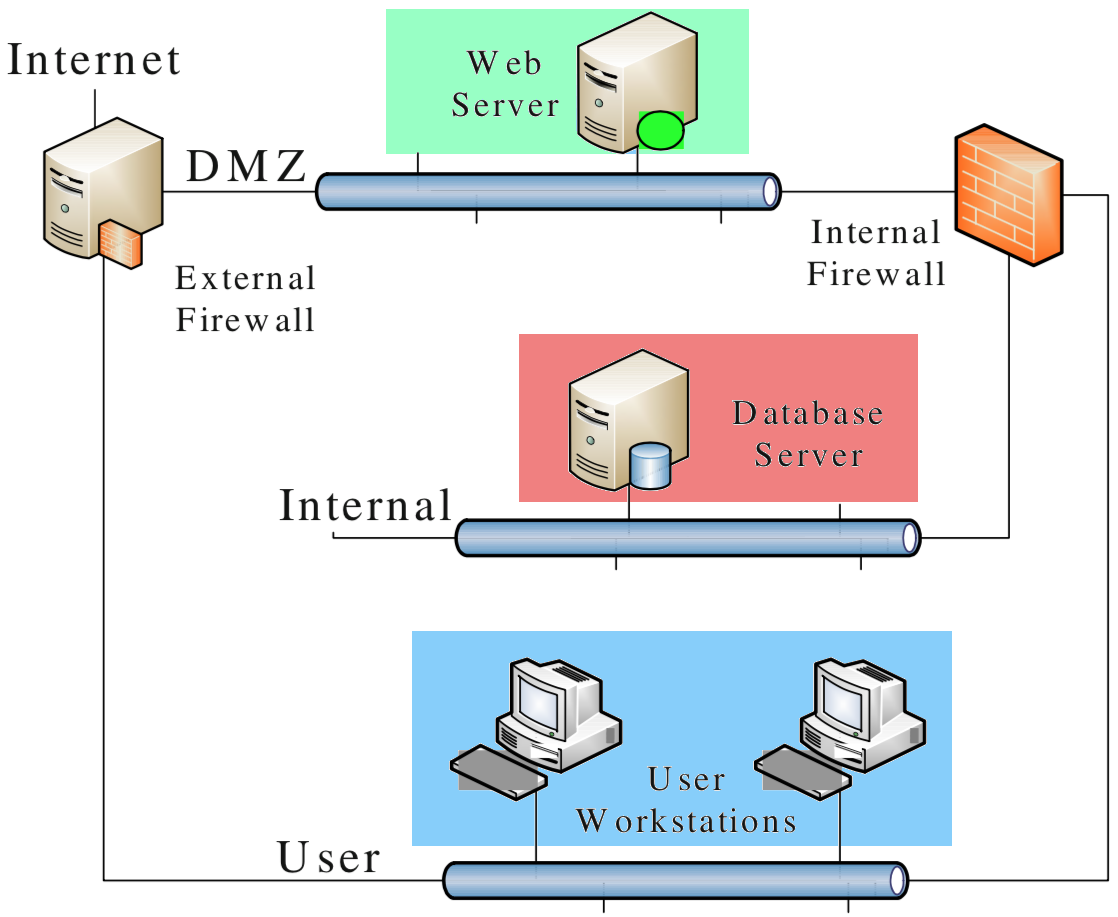}
	\caption{Network architecture used as a running example.}
	\label{egnet}
\end{figure}

As a motivating example for this work, we will consider a network architecture that could be used in a small enterprise situation. It is an example that has been used in the literature \cite{Ou2011,homer2013aggregating}. The architecture is shown in Figure \ref{egnet}. The network comprises of a Database server on the internal network. This can be accessed through an internal firewall by both the user Workstations and the Webserver, that exists in the demilitarized zone subnet. This Webserver connects to the Internet via an external firewall.

\begin{figure*}
    \centering
    \includegraphics[width=\linewidth]{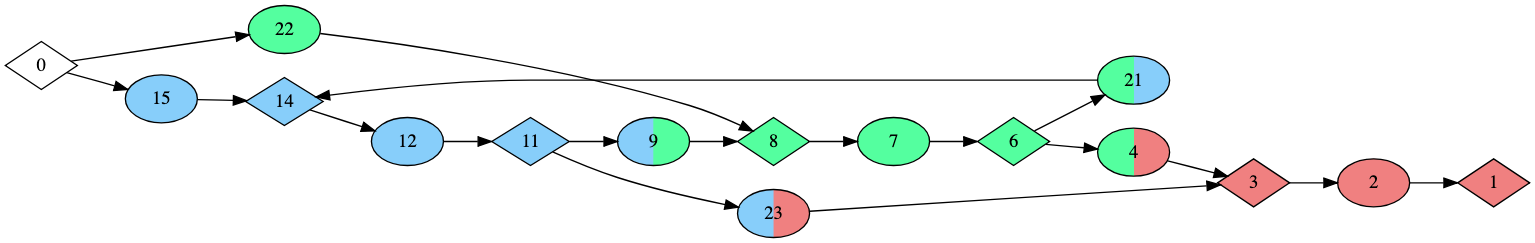}
    \caption{An excerpt of the BAG of the running example. Node colours correspond with the components of the network presented in Figure~\ref{egnet}.}
    \label{egbag}
\end{figure*}

For this scenario, we suppose a vulnerability scan has been run on the network, revealing three vulnerabilities that are present. There is a MySQL vulnerability on the Database server, an Apache vulnerability on the  Webserver, and an Internet Explorer vulnerability on the Workstations. The workstations are modelled as a single host and imagined to be all similarly set up and patched. This is not necessarily the case however, and will depend on the specific setup for the organization.

Figure~\ref{egbag} is the corresponding BAG for this situation. It is comprised of two node types; diamonds (OR nodes) represent a state that an attacker can be in like a certain level of privilege with respect to a host, and ellipses (AND nodes) are actions like exploiting a vulnerability or connecting to a host. Section \ref{sec:form} gives a formal definition of these nodes. For clarity, all Leaf nodes are removed so that the different routes to the Database server are easier to see, and have been numbered so they may be referred to (a description of each node can be found in Appendix~\ref{app:eg}). The directed edges are dependencies, an OR node can be reached if any of the parents are reached, whereas an AND node can only be reached if every parent has been reached. The colours correspond to the colours in Figure~\ref{egnet}; blue is a node that relates to the Workstations, green to the Webserver, and red to the Database server. Node 0 in white represents an attacker from the Internet.

A feature of this graph is the presence of \emph{loops} that has been already studied in the literature. For instance, the sequence of nodes $(11, 9, 8, 7, 6, 4, 3, 23, 11)$ forms a loop (cf. Definition~\ref{def:loop_cycle}).
Another important aspect of the graph is the presence of \emph{cycles}. For instance, the sequence of nodes $(14,12,11, 9, 8, 7, 6, 21, 14)$ forms a cycle due to node 21 joining back to node 14.
This cycle represents the fact that there are two different ways to gain access to a Workstation, and that once access is achieved a future state also allows a Workstation to be compromised. A user can simply access a malicious website, shown on the route along nodes 15 and 14, or alternatively an attacker could exploit the vulnerability on the Webserver (nodes 22, 8, 7, 6) and then compromise a website from there using the Webserver to gain access to the workstation.

Because this cycle can be entered into from multiple nodes (either 14 or 8), calculating the probability of an attacker reaching node 14, or indeed any other node in the cycle, cannot be done using basic approaches to solving BAGs. The \emph{monotonicity principle}, which states that an attacker will always increase their privilege \cite{Ammann2002:SGN:586110.586140}, cannot be used to remove the edge (between 21 and 14) either. This is because it is possible for an attacker to reach node 14 for the first time using the edge (21 to 14) that needs to be removed, if they enter the cycle travelling from node 22 to node 8. As such there is no loss of privilege and the graph cannot be simplified. Our approach performs the computation on the graph without the need for any simplification.


A more complete discussion of the scenario and attack graph, including the specific vulnerabilities, can be found in Appendix~\ref{app:eg}.

\subsection{BAG Formalisms}
\label{sec:form}
In this section, we introduce two different attack graph formalisms widely used in the literature for modeling a network from the security perspective. The first one is proposed in \cite{Swiler2001} and the second one in \cite{Ou2006:SAA:1180405.1180446}. We will then show in Section \ref{sec:equivalence} that the second formalism is more general than the first one and work with that representation after this section.
Note that both formalisms require the attack graph to be acyclic. We use these formalisms as a basis for generalising them to cyclic graphs. We first define cycles and loops in directed graphs.
\begin{definition}
\label{def:loop_cycle}
Given a directed graph $G = (V,E)$ with the set of nodes $V$ and the set of edges $E\subset V\times V$, a \emph{cycle} is a sequence of nodes $(v_1,v_2,\ldots, v_n)$ such that $(v_i,v_{i+1})\in E$ for all $i$ and $v_n=v_1$.
The graph is called acyclic if it does not have any cycles. A \emph{loop} is a sequence of nodes $(v_1,v_2,\ldots, v_n)$ such that $v_n=v_1$ and for any $i$, either $(v_i,v_{i+1})\in E$ or $(v_{i+1},v_{i})\in E$.
\end{definition}
Loops can be seen as cycles in the undirected version of the graph, i.e., when the pair $(v,v')$ is treated the same as $(v',v)$.
According to Definition~\ref{def:loop_cycle}, any cycle is also a loop but in the sequel, we use the word `loop' to refer to those that are not cycles.
Moreover, an acyclic graph can still have loops. Most of the literature on BAGs is focused on acyclic graphs as defined next.

\subsubsection{Plain BAGs}

\begin{definition}
\label{def:BAG1}
An attack graph $G$ is a directed graph $G =(E\cup C, R_r\cup R_i)$ where $E$ is a set of exploits, $C$ a set of conditions, and $R_r\subseteq C\times E$ and $R_i\subseteq E\times C$.
\end{definition}
\begin{remark}
According to Definition~\ref{def:BAG1}, the attack graph is \emph{bipartite}, i.e., the set of nodes of the graph is divided into two disjoint and independent sets $E$ and $C$ such that every edge can only connect a node from one to another. That is why the set of nodes is partitioned into a subset of $C\times E$ and a subset of $E\times C$.
\end{remark}
The edges connecting conditions to exploits have a particular meaning: all the conditions connected to an exploit must be satisfied in order to execute that exploit. This is the equivalent of taking the conjunction of the incoming conditions to the exploit. Similarly, any of the exploits connected to a condition can be used to satisfy that condition. This is equivalent to a disjunction between multiple exploits that satisfy the same condition. Such an interpretation together with individual scores assigned to the nodes fully characterises the attack model.
\begin{definition}
\label{def:score1}
Given an \emph{acyclic} attack graph $G =(E\cup C, R_r\cup R_i)$, and an individual score assignment function $p : E \cup C \rightarrow [0,1]$, the cumulative score function $P : E\cup C \rightarrow[0,1]$ is defined as
\begin{align}
P(e) &= p(e)\cdot\Pi_{c\in R_r(e)}P(c)\nonumber\\
P(c) &= p(c),\quad \text{if } R_i(c) = \emptyset\label{eq:cumulative}\\
P(c) &= p(c)\cdot\oplus_{e\in R_i(c)} P(e),\quad \text{if } R_i(c) \neq \emptyset,\nonumber
\end{align}
where $\oplus_{e\in R_i(c)} P(e)$ is the probability of the union of exploits in $R_i(c)$ and is computed assuming the exploits are independent.   
\end{definition}
The acyclic attack graph $G =(E\cup C, R_r\cup R_i)$ together with the individual scores defines a \emph{plain BAG}.
Note that attack graphs can in general have cycles but plain BAGs are defined with acyclic attack graphs.

\subsubsection{AND/OR BAGs}

The second formalism of BAG is defined by Ou \cite{Ou2006:SAA:1180405.1180446} and is used in MulVAL \cite{Ou2005:MLN:1251398.1251406}.
\begin{definition}
\label{def:BAG2}
A Bayesian attack graph is defined as a directed acyclic graph $\mathcal G = (\mathcal V,\mathcal E)$ where nodes $\mathcal V$ are connected by edges $\mathcal E$. The set of nodes is comprised of three types of nodes, $\mathcal V = V_l \cup V_a \cup V_o$, and edges are defined as $e_{ij} \in E, e_{ij}=e(v_i,v_j)$ where each edge $e$ defines a mapping from node $v_i$ to node $v_j$.
\end{definition}

The sets of nodes are defined as follows:
\begin{itemize}
\item
$V_l$: the {\em leaves} in the graph, having no parents, and represent specific configurations and conditions in the network; this includes information about programs running on a host, network connection information in the form of HACLs (host access control lists), and the existence of vulnerabilities.
\item
$V_a$: the AND nodes, which have requisite conditions {\em all} of which have to be fulfilled in order to be accessed. In other words there is a conjunctive relationship between the parents of such a node. This set of nodes is used to represent specific actions that can be taken when the conditions are fulfilled; this can be something like movement between hosts when an attacker has fulfilled the prerequisite of access to one machine and there exists a configuration node for access between the two nodes, or could be the remote exploit of a specific vulnerability given remote access and the existence of the vulnerability as prerequisites.
\item
$V_o$: the OR nodes, which have requisite conditions of which {\em at least one} must be fulfilled in order to be accessed. There is a disjunction between the parents of such a node. These are specific micro-states in the network that define something about an attacker's position in the system, for example the ability to execute arbitrary code on a specific host or network access to a specific host. A macro-state for an attacker would be an enumeration of these nodes demonstrating the privilege they have with respect to every host on the network.
\end{itemize}


Vulnerabilities in the network have a chance to be exploited when their preconditions are fulfilled, and by exploiting a vulnerability an attacker achieves a specific state. This state, once reached, may then afford the attacker a privilege level on the network that is a requirement for another exploit, or node. A chain of nodes in the network connected in this way represents an attack path or route. 

We define the \emph{access probability} $P(v)$ on the node $v$ as the likelihood of the node being reached in an attack situation.
\begin{definition}
\label{def:score2}
For a given BAG $\mathcal G = (\mathcal V,\mathcal E)$ and a \emph{local probability} function $p:\mathcal V\rightarrow[0,1]$,
the access probability $P:\mathcal V\rightarrow[0,1]$ is defined recursively using the access probabilities of all parents to the node in conjunction with the \emph{local probability} by
\begin{equation}
\label{eq:scores}
P(v) = \begin{cases}
p(v) &\mbox{if } v \in V_l\\
p(v)\displaystyle\prod_{v'\in pa(v)} P(v') &\mbox{if } v \in V_a\\
p(v)\big[1-\displaystyle\prod_{v'\in pa(v)} (1-P(v'))\big] &\mbox{if } v \in V_o
\end{cases}
\end{equation}

where $pa(v)$ represents the parent set of the node $v\in\mathcal V$, $pa(v) := \{v'\in\mathcal V|(v',v)\in\mathcal E\}$.

\end{definition}
The access probability has a slightly different interpretation depending on the specifics of the node.
For $v \in V_o$,  $P(v)$ represents the probability that the attacker will achieve the state described by node $v$.
For $v \in V_a$,  $P(v)$ represents the probability that an attacker will travel along that specific route to reach the goal state that follows.
For $v \in V_l$,  $P(v)$ represents the probability of successful exploitation if the node defines a vulnerability, or is the probability that a specific entry-route will be used.

\begin{remark}
\label{rem:indep}
Access probabilities $P$ defined in \eqref{eq:scores} assume that probabilities $P(v_{pa})$ are independent from each other and takes the product of these probabilities to find the access probability for their child node. This assumption is only true if the graph of the BAG does not have loops. Otherwise, $P(v)$ in \eqref{eq:scores} will only be an approximation of true access probabilities that can be computed using joint distributions to reflect the dependencies between the related events. One of these exact methods is Variable Elimination discussed in Section~\ref{sec:VE}. 
\end{remark}



\subsection{Relation Between the Two Formalisms}
\label{sec:equivalence}
In the following proposition, we show that the AND/OR definition of BAGs is more general than plain BAGs, as it abstracts away the type of nodes being exploits or conditions.  Instead, it puts emphasis on their role in the computation of access probabilities.
\begin{proposition}
Any plain BAG modelled as in Definitions~\ref{def:BAG1}-\ref{def:score1} can be transformed into a BAG modelled as in Definitions~\ref{def:BAG2}-\ref{def:score2}. 
\end{proposition}
\begin{IEEEproof}
Suppose we have a plain BAG with the acyclic attack graph $G =(E\cup C, R_r\cup R_i)$. Define the set of leaf nodes as $V_l := \{c\in C \,|\, R_i(c) = \emptyset\}$, the set of OR nodes $V_o := C\backslash V_l$, and the set of AND nodes $V_a := E$. Take $\mathcal V = V_l \cup V_a \cup V_o$ and $\mathcal E = R_r\cup R_i$. Then $\mathcal G = (\mathcal V,\mathcal E)$ is an attack graph satisfying all the requirements of Definition~\ref{def:BAG2}.
Note that attack graphs of AND/OR BAGs are not necessarily bipartite, which makes them more general than plain BAGs.
\end{IEEEproof}

\section{Computation of Access Probabilities}
\label{sec:access}
The main approach for computing access probabilities of all nodes is to translate the model into a Bayesian network (BN) and apply off-the-shelf techniques developed in the literature for BNs. We first provide the translation of the BAG into a BN in section~\ref{sec:BN} and then discuss \emph{variable elimination} as one of the techniques for performing probability computations over BNs in section~\ref{sec:VE}.

\subsection{BAG Translation to a Bayesian Network}
\label{sec:BN}

\begin{definition}
\label{def:BN}
A Bayesian network (BN) is a tuple $\mathfrak B = (\mathcal V,\mathcal E,\mathcal T)$. The pair 
$(\mathcal V,\mathcal E)$ is a directed acyclic graph representing the structure of the network. 
The nodes in $\mathcal V$ are (discrete or continuous) random variables and the arcs in $\mathcal E$ represent the dependence relationships among the random variables. 
The set $\mathcal T$ contains conditional probability distributions (CPD) in forms of tables or density functions for discrete and continuous random variables, respectively.
\end{definition}

In a BN, knowledge is represented in two ways: 
qualitatively, as dependencies between variables by means of a directed acyclic graph; and 
quantitatively, as conditional probability distributions attached to the dependence relationships. 
Each random variable $v_i\in\mathcal V$ is associated with a conditional probability distribution $\text{Prob}(v_i|pa(v_i))$.

\begin{proposition}
\label{prop:BN}
Any BAG $\mathcal G = (\mathcal V,\mathcal E)$ as in Definition~\ref{def:BAG2} with local probability function $p:\mathcal V\rightarrow [0,1]$ in Definition~\ref{def:score2} can be translated into a BN $\mathfrak B = (\mathcal V,\mathcal E,\mathcal T)$. The random variables in $\mathcal V$ are all Boolean and the probability tables in $\mathcal T$ are constructed as follows. For all $v\in V_l$,
\begin{equation}
\label{eq:BN_leaf}
\text{Prob}(v=1) = p(v) \quad \text{ and }\quad  \text{Prob}(v=0) = 1-p(v).
\end{equation}
For all $v\in V_a$, let $pa(v)=\mathbf{1}$ indicate that all variables in $pa(v)$ take value equal to one. Then,
\begin{equation}
\label{eq:BN_and}
\begin{cases}
\text{Prob}(v=1|pa(v)=\mathbf{1}) = p(v),\\
\text{Prob}(v=1|pa(v)\neq\mathbf{1}) = 0,\\
\text{Prob}(v=0|pa(v)=\mathbf{1}) = 1-p(v),\\
\text{Prob}(v=0|pa(v)\neq\mathbf{1}) = 1.
\end{cases}
\end{equation}
For all $v\in V_o$, let $pa(v)=\mathbf{0}$ indicate that all variables in $pa(v)$ take value equal to zero. Then,
\begin{equation}
\label{eq:BN_or}
\begin{cases}
\text{Prob}(v=1|pa(v)=\mathbf{0}) = 0,\\
\text{Prob}(v=0|pa(v)=\mathbf{0}) = 1\\
\text{Prob}(v=1|pa(v)\neq\mathbf{0}) = p(v),\\
\text{Prob}(v=0|pa(v)\neq\mathbf{0}) = 1-p(v).
\end{cases}
\end{equation}
Then if the BAG $\mathcal G$ does not have any loops, we get $P(v) = \text{Prob}(v=1)$ for all $v\in\mathcal V$ with access probabilities $P$ defined in \eqref{eq:scores}.
\end{proposition}

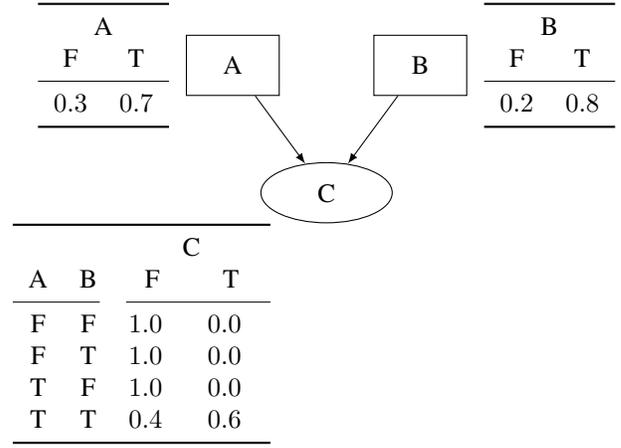
\begin{figure}
    \centering
    \begin{tikzpicture}[node distance=1cm and 0cm]
    \node[draw, minimum size=0.8cm, text width=1cm, align=center] (A) {A};
    \node[draw, ellipse, minimum size=0.8cm, below right=of A, text width=1cm, align=center] (C) {C};
    \node[draw, minimum size=0.8cm, above right=of C, text width=1cm, align=center] (B) {B};
    \path (A) edge[-latex] (C)
    (B) edge[-latex] (C);
    \node[right=0.1cm of B]
        {
        \begin{tabular}{M{1}M{1}}
        \toprule
        \multicolumn{2}{c}{B} \\
        \multicolumn{1}{c}{F} & \multicolumn{1}{c}{T} \\
        \cmidrule{1-2}
        0.2 & 0.8 \\
        \bottomrule
        \end{tabular}
        };
    \node[left=0.1cm of A]
        {
        \begin{tabular}{M{1}M{1}}
        \toprule
        \multicolumn{2}{c}{A} \\
        \multicolumn{1}{c}{F} & \multicolumn{1}{c}{T} \\
        \cmidrule{1-2}
        0.3 & 0.7 \\
        \bottomrule
        \end{tabular}
        };
    \node[below left=0.0cm of C]
        {
        \begin{tabular}{ccM{2}M{2}}
        \toprule
        & & \multicolumn{2}{c}{C} \\
        \multicolumn{1}{l}{A} & \multicolumn{1}{l}{B} & \multicolumn{1}{c}{F} & \multicolumn{1}{c}{T} \\
        \cmidrule(r){1-2}\cmidrule(l){3-4}
        F & F & 1.0 & 0.0 \\
        F & T & 1.0 & 0.0 \\
        T & F & 1.0 & 0.0 \\
        T & T & 0.4 & 0.6 \\
        \bottomrule
        \end{tabular}
        };
    \end{tikzpicture}
    \caption{A simple BAG with the associated probability tables constructed according to Proposition~\ref{prop:BN}. Local probabilities are $p(A) = 0.7$, $p(B) = 0.8$, and $p(C) = 0.6$.}
    \label{fig:ve_eg}
\end{figure}

Figure~\ref{fig:ve_eg} illustrates the construction of probability tables for an AND node. In this figure, the local probabilities are $p(A) = 0.7,p(B) = 0.8$, and $p(C) = 0.6$. The probability tables for $A$ and $B$ are constructed according to \eqref{eq:BN_leaf} and for $C$ according to \eqref{eq:BN_and}.

\subsection{Variable Elimination}
\label{sec:VE}
Based on the results of section~\ref{sec:BN}, the computation of access probabilities is equivalent to the computation of marginal probabilities $\text{Prob}(v=1)$ in the associated BN. \emph{Variable elimination} (VE) is a simple and general algorithm developed in the literature that computes exact values of these marginal probabilities (e.g., \cite{Koller:2009}). Given that the structure of the graph $(\mathcal V,\mathcal E)$ models the independence between random variables associated with the nodes, we can obtain the joint distribution of the variables as the product of the conditional probability tables $\prod_{v'\in\mathcal V}\text{Prob}(v'|pa(v'))$. Then the marginal probabilities are computed by taking the sum over the unwanted variables:
\begin{equation}
\label{eq:sum_prod}
    \text{Prob}(v) = \sum_{v'\neq v, v'\in\{0,1\}}\prod_{v'\in\mathcal V}\text{Prob}(v'|pa(v')).
\end{equation}
VE provides a procedure for the computation of the sum-product in \eqref{eq:sum_prod}. The main goal of the algorithm is to specify at each iteration which tables to multiply and which variable to sum over. The reader may refer to the book \cite[Chapter 9]{Koller:2009} for a detailed discussion on VE. 

\begin{proposition}
The access probabilities of Definition~\ref{def:score2} are equal to the marginal probabilities of the BN of Definition~\ref{def:BN} obtained by variable elimination in the case that the BAG does not have any loops.  
\end{proposition}


\begin{figure}
    \centering
    \includegraphics[width=\linewidth]{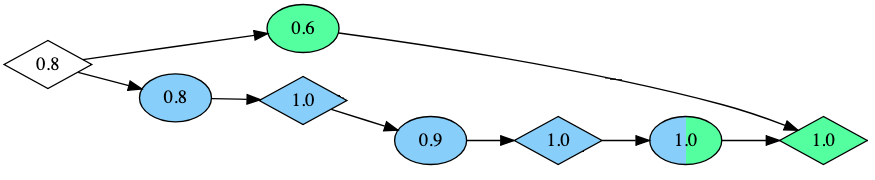}
    \caption{A subgraph of the running example indicating a loop.}
    \label{fig:ve}
\end{figure}

In the case that the BAG does have loops, the access probabilities will not necessarily be equal to the marginal probabilities calculated using VE. This can be shown by taking the first loop from Figure \ref{egbag} and using the local probabilities shown in Figure \ref{fig:ve}, then marginalizing node 8 with VE and calculating the access probability for the same node using \eqref{eq:scores} of Definition~\ref{def:score2}. The former gives a probability of $0.7104$ for node 8 being True; the latter gives $0.7795$. This discrepancy is due to nodes 15 and 22 being assumed to be independent while having a common ancestor (cf. Remark~\ref{rem:indep}). In other words the probability at node 0 is being allowed to contribute more than it should to the final result, hence the higher calculated probability.
This level of discrepancy is already studied in the literature: VE is known to give the exact values while other computational approaches give approximate values for BAGs with loops \cite{Munoz-Gonzalez2016}. 

\comment{
\begin{table} 
    \caption{Elimination for Figure \ref{fig:ve_eg}}
    \label{tab:vefactors}
    \begin{tabular}{cccc}
        \toprule
	    Variable & Factor Form & Variables  & New Factor\\  
        Eliminated &  & Involved & \\
        \midrule
	    A & $\sum_{A}p(C|A,B)p(A)$ & A, B, C & $\tau_{1}(C,B)$\\
	    B & $\sum_{B}\tau_{1}(C,B)p(B)$ & B, C & $\tau_{2}(C)$\\
	    C & $\sum_{C}\tau_{2}(C)p(D|C)$ & C & $\tau_{3}$\\
        \bottomrule
    \end{tabular}
\end{table}

This process of calculating the probabilities is analogous to marginalizing all other probabilities at each node. This can be done via the process of variable elimination to demonstrate the correctness of this approach. Using the  simple network from Figure \ref{fig:ve_eg} we can eliminate all other variables using the factors in table~\ref{tab:vefactors} in order to calculate the marginal probability at node D. Thus we end with the result that the marginal probability for D to be true is 0.56; note that this is the same as the result that would be achieved by simply using the conjunctive and disjunctive formulae to the graph according to the node types.

Due to the nature of this formulation of attack graphs, with probabilities less than 1 being assigned to leaf nodes, the internal portion of a graph is always comprised of probability tables that either represent conjunction or disjunction and as such the formula for each of these can be easily applied to calculate the marginal probabilities and achieve the same result as variable elimination.
}

\section{Combinational Circuits with Probabilistic Inputs}
\label{sec:combinational}
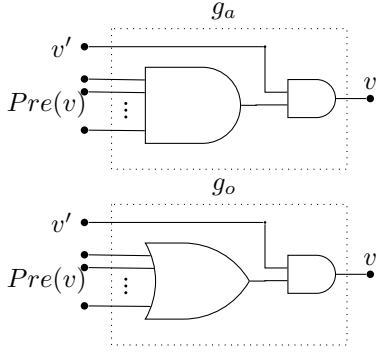
\begin{figure}
    \centering
    \begin{tikzpicture}
    \node (vin) at (0, 0.6) [circle,fill,inner sep=1pt]{};
    \node (vin2) at (2.4, 0.6) [circle,fill,outer sep=0pt, inner sep=0pt] {};
    \node (in1) at (0, 0.17) [circle,fill,inner sep=1pt]{};
    \node (prev) at (0, 0) [circle,fill,inner sep=1pt]{};
    \node (in2) at (0, -0.51) [circle,fill,inner sep=1pt]{};
    \node (vinl) at ($(vin) + (-0.3,0)$) {$v'$};
    \node (prevl) at ($(prev) + (-0.5,-0.17)$) {$Pre(v)$};
    
    \node (dot1) at ($(prev) + (0.55,-0.155)$) [circle,fill,inner sep=0.4pt]{};
    \node (dot2) at ($(prev) + (0.55,-0.245)$) [circle,fill,inner sep=0.4pt]{};
    \node (dot3) at ($(prev) + (0.55,-0.335)$) [circle,fill,inner sep=0.4pt]{};
    
    \node (end) at (3.8, -0.09) [circle,fill,inner sep=1pt]{};

    \node[and gate US, draw, rotate=0, logic gate inputs=nnnnn] at ($(prev) + (1.4, -0.18)$) (and) {}; 
    \node[and gate US, draw, rotate=0, logic gate inputs=nn] at ($(prev) + (3, -0.09)$) (and2) {}; 

    \draw (in1) -- (and.input 1);
    \draw (prev) -- (and.input 2);
    \draw (in2) -- (and.input 5);
    
    \draw (vin) -- (vin2);
    \draw (vin2) |- (and2.input 1);
    \draw (and.output) -- ([xshift=0.2cm]and.output) |- (and2.input 2);
    \draw (and2.output) -- (end) node[above] {$v$};
    \draw[thin,dotted]     ($(vin.north west)+(0.4,0.2)$) node[above, xshift=1.5cm] {$g_a$} rectangle ($(end.south east)+(-0.35,-0.9)$);

\end{tikzpicture}
    \begin{tikzpicture}
    \node (vin) at (0, 0.6) [circle,fill,inner sep=1pt]{};
    \node (vin2) at (2.4, 0.6) [circle,fill,inner sep=0pt] {};
    \node (in1) at (0, 0.17) [circle,fill,inner sep=1pt]{};
    \node (prev) at (0, 0) [circle,fill,inner sep=1pt]{};
    \node (in2) at (0, -0.51) [circle,fill,inner sep=1pt]{};
    \node (vinl) at ($(vin) + (-0.3,0)$) {$v'$};
    \node (prevl) at ($(prev) + (-0.5,-0.17)$) {$Pre(v)$};
    
    \node (dot1) at ($(prev) + (0.55,-0.155)$) [circle,fill,inner sep=0.4pt]{};
    \node (dot2) at ($(prev) + (0.55,-0.245)$) [circle,fill,inner sep=0.4pt]{};
    \node (dot3) at ($(prev) + (0.55,-0.335)$) [circle,fill,inner sep=0.4pt]{};
    
    \node (end) at (3.8, -0.09) [circle,fill,inner sep=1pt]{};

    \node[or gate US, draw, rotate=0, logic gate inputs=nnnnn] at ($(prev) + (1.4, -0.18)$) (and) {}; 
    \node[and gate US, draw, rotate=0, logic gate inputs=nn] at ($(prev) + (3, -0.09)$) (and2) {}; 

    \draw (in1) -- (and.input 1);
    \draw (prev) -- (and.input 2);
    \draw (in2) -- (and.input 5);
    
    \draw (vin) -- (vin2);
    \draw (vin2) |- (and2.input 1);
    \draw (and.output) -- ([xshift=0.2cm]and.output) |- (and2.input 2);
    \draw (and2.output) -- (end) node[above] {$v$};
    \draw[thin,dotted]     ($(vin.north west)+(0.4,0.2)$) node[above, xshift=1.5cm] {$g_o$} rectangle ($(end.south east)+(-0.35,-0.9)$);

\end{tikzpicture}
    \caption{Logic gate representation of AND and OR nodes.}
    \label{fig:logic}
\end{figure}


As one of the main contributions of this paper, we look at the attack graph from a different perspective and model it as a deterministic combinational circuit with probabilistic inputs. This interpretation paves the way towards including and analysing cycles directly in the computation of access probabilities over attack graphs.
Combinational circuits are mainly studied in the literature \cite{5989992,5361265} from the perspective of constructing a certain distribution on the output of the circuit by applying random inputs.

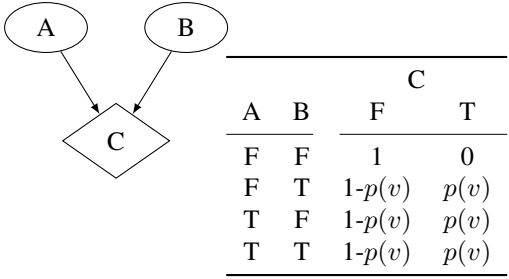
\begin{figure}
    \centering
    \begin{tikzpicture}[node distance=1cm and 0cm]
    \node[draw, ellipse, minimum width=1.1cm, align=center] (A) {A};
    \node[draw, ellipse, minimum width=1.1cm, align=center, right= 0.75cm of A] (B) {B};
    \path (A) -- (B) coordinate[midway] (mid);
    \node[draw, diamond, aspect=2, minimum size=1cm, below = of mid, align=center] (C) {C};
    \path (A) edge[-latex] (C);
    \path (B) edge[-latex] (C);
    \node[below right=0.02cm of B]
        {
        \begin{tabular}{cccc}
        \toprule
        && \multicolumn{2}{c}{C} \\
        A & B & \multicolumn{1}{c}{F} & \multicolumn{1}{c}{T} \\
        \cmidrule(r){1-2}\cmidrule(l){3-4}
        F & F & 1 & 0 \\
        F & T & 1-$p(v)$ & $p(v)$ \\
        T & F & 1-$p(v)$ & $p(v)$ \\
        T & T & 1-$p(v)$ & $p(v)$ \\
        \bottomrule
        \end{tabular}
        };
    \end{tikzpicture}
    \caption{A graph using definitions~\ref{def:BAG2}-\ref{def:score2}, with local probability in OR node C.}
    \label{fig:precon}
\end{figure}

\begin{figure}
    \centering
    \begin{tikzpicture}[node distance=1cm and 0cm]
    \node[draw, ellipse, minimum width=1.1cm, align=center] (A) {A};
    \node[draw, ellipse, minimum width=1.1cm, align=center, right= 0.75cm of A] (B) {B};
    \path (A) -- (B) coordinate[midway] (mid);
    \node[draw, diamond, aspect=2, minimum size=1cm, below = of mid, align=center] (C1) {C$^\prime$};
    \node[draw, minimum size=0.8cm, align=center, right=0.75cm of C1] (v) {$v'$};
    \path (C1) -- (v) coordinate[midway] (mid1);
    \node[draw, ellipse, minimum width=1.1cm, align=center, below= 1cm of mid1] (C) {C};
    \path (A) edge[-latex] (C1);
    \path (B) edge[-latex] (C1);
    \path (v) edge[-latex] (C);
    \path (C1) edge[-latex] (C);
    \node[right=0cm of v]
        {
        \begin{tabular}{cc}
        \toprule
        \multicolumn{2}{c}{$v'$} \\
        \multicolumn{1}{c}{F} & \multicolumn{1}{c}{T} \\
        \cmidrule(l){1-2}
        1-$p(v)$ & $p(v)$ \\
        \bottomrule
        \end{tabular}
        };
    \node[below left=0.08cm of C1]
        {
        \begin{tabular}{cccc}
        \toprule
        && \multicolumn{2}{c}{C$^\prime$} \\
        A & B & \multicolumn{1}{c}{F} & \multicolumn{1}{c}{T} \\
        \cmidrule(r){1-2}\cmidrule(l){3-4}
        F & F & 1 & 0 \\
        F & T & 0 & 1 \\
        T & F & 0 & 1 \\
        T & T & 0 & 1 \\
        \bottomrule
        \end{tabular}
        };
    \node[below=0.15cm of C]
        {
        \begin{tabular}{cccc}
        \toprule
        && \multicolumn{2}{c}{C} \\
        C$'$ & $v'$ & \multicolumn{1}{c}{F} & \multicolumn{1}{c}{T} \\
        \cmidrule(r){1-2}\cmidrule(l){3-4}
        F & F & 1 & 0 \\
        F & T & 1 & 0 \\
        T & F & 1 & 0 \\
        T & T & 0 & 1 \\
        \bottomrule
        \end{tabular}
        };
    \end{tikzpicture}
    \caption{The graph of Figure~\ref{fig:precon} transformed into a graph with probabilities only on leaves.}
    \label{fig:postcon}
\end{figure}
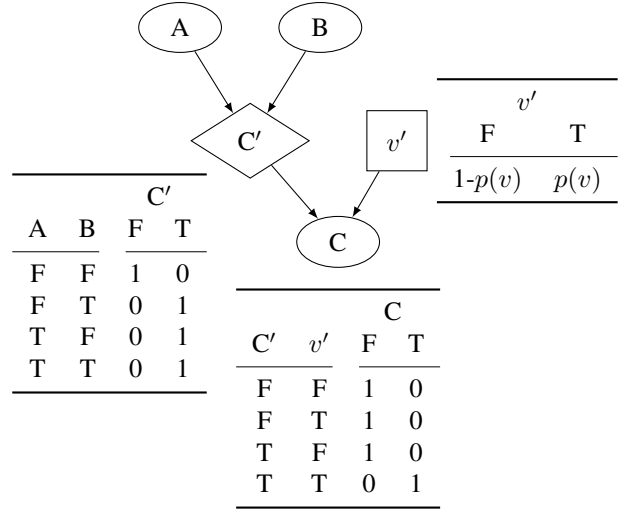
Let us enlarge the attack graph $(\mathcal V,\mathcal E)$ with the set of nodes $\mathcal V = \{v_1,v_2,\ldots, v_n\}$ to an augmented attack graph $(\bar{\mathcal V},\bar{\mathcal E})$ with nodes $\bar{\mathcal V} := \{v_1,v_2,\ldots, v_n,v'_1,v'_2,\ldots,v'_n\}$ and edges $\bar{\mathcal E}:= \mathcal E\cup\{(v'_1,v_1),\ldots,(v'_n,v_n)\}$. The augmented graph is obtained by adding one node $v'$ for each node $v\in \mathcal V$ and connecting it directly to $v$. The added node $v'$ has the role of modeling local probabilities at node $v$ and renders the behaviour of this node non-probabilistic. Assuming a delay in the computation of the value of the node, we get one of the following two equations for each node:
\begin{align*}
    &v(k+1) = g_a(pa(v)(k),v')\\
    &\quad\quad\quad\quad\text{ or }\\
    &v(k+1) = g_o(pa(v)(k),v'),
\end{align*}
where $v(k+1)$ and $pa(v)(k)$ indicate respectively the value of node $v$ at $k+1^{\text{st}}$ iteration and the values of the parent nodes of $V$ at $k^{\text{th}}$ iteration.
The two functions $g_a$ and $g_o$ correspond to the \emph{AND} node or the \emph{OR} node respectively, and are depicted in Figure~\ref{fig:logic} using logic gates. Note that the Leaf nodes can be treated as either AND nodes or OR nodes.

A demonstration of enlarging an attack graph to make internal nodes behave in a non-probabilistic way can be seen in Figures \ref{fig:precon} and \ref{fig:postcon}. Figure \ref{fig:precon} is an AND/OR BAG, with the probability table for the OR node shown to the right. We can move the local probability to a Leaf node, and as such all the internal probability tables simply become logical AND and OR tables. This is shown in Figure~\ref{fig:postcon}, where the equivalent BAG is shown, and the left and central probability tables can be recognised as logical OR and AND truth tables respectively, with the rightmost probability table containing the local probability on a leaf of the graph.

\begin{theorem}
The behaviour of an attack network can be modelled as a combinational circuit with probabilistic inputs. The value of the variables are changing according to the equation
\begin{equation}
\label{eq:dynamical_logic}
\left\{
    \begin{array}{l}
       v_1(k+1) = g_1(pa(v_1)(k),v_1') \\
        v_2(k+1) = g_2(pa(v_2)(k),v_2')\\
        \quad \vdots\\
        v_n(k+1) = g_n(pa(v_n)(k),v_n'),
    \end{array}
    \right.
\end{equation}
where $k=0,1,2,\ldots$ models the progression of an attacker in gaining access to the nodes or satisfying conditions along the time axis, $g_i\in\{g_a,g_o\}$ for all $i$, with $g_a,g_o$ defined according to Figure~\ref{fig:logic}. The nodes $v'_i$ take values $\{0,1\}$ according to the local probabilities. The notation $(k)$ is used for the $k^{\text{th}}$ iteration.
\end{theorem}

Access probabilities are equivalent to the computation of reachability probabilities over the augmented graph:
\begin{equation}
\label{eq:reach}
    \text{Prob}(v=1) = \text{Prob}\{v(k) = 1 \text{ for some } k\},
\end{equation}
where the probability is computed over all combination of values of $\{v_1',v_2',\ldots,v_n'\}\in\{0,1\}^n$. Unlike previous formalisms that are not able to handle cycles, our new interpretation can easily encode cycles without the need for any modification.

Next we prove a property of this interpretation that helps in developing our algorithm for the computation of reachability probabilities in \eqref{eq:reach}.

\begin{proposition}
The system of equations \eqref{eq:dynamical_logic} is monotonically increasing, i.e., $v_i(k+1)\ge v_i(k)$ for any $k$ and $i$ and any instantiation of $\{v'_1,v'_2,\ldots, v'_n\}$.
\end{proposition}
\begin{IEEEproof}
Fix an instantiation of $\{v'_1,v'_2,\ldots, v'_n\}$.
First we show that the function $g_a$ is monotonically increasing, which is the property that $g_a(w,v')\geq g_a(w',v')$ for any $w,w'$ with $w\ge w'$ element-wise.
Note that $g_a(w',v')=0$ if an element of $w'$ or $v'$ is zero, which means the inequality holds. If all elements of $w'$ and $v'$ are one, then all elements of $w$ is also one, which means $g_a(w,v')=g_a(w',v')=1$ and the inequality holds.\\
Next we show that the function $g_o$ is monotonically increasing, which is the property that $g_o(w,v')\geq g_o(w',v')$ for any $w,w'$ with $w\ge w'$ element-wise.
This holds due the identity $g_o(w,v') = 1- g_a(1-w,1- v')$ that the OR gate is the complement of the AND gate:\\
$w\ge w'\Rightarrow 1-w\le 1-w'$\\
$\Rightarrow g_a(1-w,1- v')\le g_a(1-w',1- v')$\\
$\Rightarrow 1-g_a(1-w,1- v')\ge 1- g_a(1-w',1- v')$\\
$\Rightarrow g_o(w,v')\ge g_o(w',v')$.\\
Now the claim follows inductively from the fact that initially $v(k)=0$ for $k=0$, and functions $g_a$ and $g_b$ are non-negative and monotonically increasing.
\end{IEEEproof}

\begin{theorem}
\label{thm:finite_k}
The solution of \eqref{eq:dynamical_logic} converges to a unique steady state in finite time, i.e., there is a time instance $k^\ast$ such that $v_i(k^\ast+1) = v_i(k^\ast)$ for any $i$ and any instantiation of $\{v'_1,v'_2,\ldots, v'_n\}$.
\end{theorem}
\begin{theorem}
\label{thm:CCP}

For an acyclic BAG with the time instance $k^\ast$ defined in the previous theorem, $\text{Prob}(v(k^\ast)=1)$ is the same as the probability computed via Variable Elimination on the BAG. Furthermore, if the BAG does not have loops, this quantity is the same as access probabilities in \eqref{eq:scores}.
\end{theorem}

Reachability probabilities $\text{Prob}(v(k^\ast)=1)$ are well-defined on combinational circuits regardless of the existence of cycles. Therefore, the above theorem gives a nice direction for generalizing computations to cyclic BAGs. In the next section, we provide an algorithmic computation of probabilities while replacing the joint distributions with product of marginal distributions, similar to \eqref{eq:scores}.


\comment{
In order to transform a type of attack graph commonly seen in literature, formalised by Yun, Xi-shan, and Zhi-chang \cite{Yun2011}, we must separate the exploitation of the vulnerability from it's existence, and introduce the probability of that vulnerability being exploited into the leaf node that represents the vulnerabilities presence in the network. More formally, we take a Bayesian attack graph $\textbf{G} = (V,E)$, where the set of nodes is comprised of $V = V_s \cup V_e$ ($V_s$ being the set of state or condition nodes, and $V_e$ being the set of exploitations). In order to create our new graph, $G`$ we need to create a new set of nodes, $V'$, and split the exploitation set $V_e$ such that for every exploitation node $v_i \in V_e$ we add both a vulnerability in the form of a LEAF node $v_k \in V'_l$ and an exploitation node in the form of an AND node $v_i' \in V'_a$. We also transform the state nodes into OR nodes, $v_j \in V_a$ becomes $v_j' \in V'_o$ and as such the new graph's set of nodes is represented by $V' = V'_l \cup V'_a \cup V'_o$.

The set of edges $E'$ is generated from $E$, with the edge $e_{ij}$ being transformed by the mapping $e_{ij} \xrightarrow{} (e_{ki'},e_{i'j'}), e_{ij} \in E, {e_{ki'},e_{i'j'}} \in E'$. From this we can now construct our new graph as $G' = (V',E')$, and $G'$ has the property that only the LEAF nodes require defined probabilities, and all internal parts of the graph can be calculated using the conjunctive and disjunctive formulae.
}

\section{Calculation on Cyclic BAGs}
\label{sec:calculation}
\subsection{Algorithmic Inference}
We have created an algorithm to propagate probabilities through an attack graph and thus generate a BAG based on the probabilities assigned at the leaves of the graph. The algorithm, shown in Algorithm \ref{alg:main}, works by detecting all attack paths that lead to a node. It moves through each step in each attack path collapsing cycles by the method previously discussed (prevention of multiple instances of the same node from contributing to the probability multiple times) and calculating probabilities using the recursive conjunction and disjunction functions. Probabilities on a path are calculated fully for each node as the chance of a node being reached, but will not necessarily reflect its contribution to the proceeding nodes. In essence this means that when a node is being calculated, all the nodes that contribute to this probability are identified and only allowed to contribute to the final calculation once, thus ensuring that no nodes in a loop inflate the final probability by being counted multiple times. The order of calculation for the nodes does not matter and can be performed in a random order as contributions to a probability are explicitly calculated for each node every time. 


\begin{figure}[tp]
\vspace*{-\baselineskip}
 \removelatexerror
\begin{minipage}{\columnwidth}
\begin{algorithm}[H]
\removelatexerror
	\SetAlgoLined
	\KwIn{Attack Graph; nodes $v$ in $V$ with local probability $p(v)$}
	\KwOut{Bayesian Attack Graph; nodes $v$ in $V$ with access probability $P(v)$}
	
	\For{$v \in V$}{
		$P(v) = \texttt{RecursiveProbability(}v,v\texttt{)}$
	}
	\Fn{\FRecurs{node $v$, origin node $v_{origin}$}}{
		\If{$v \in V_l$}{
			\Return{$p(v)$}
		}
		\Else{
			\For{$v_{pa} \in V_{parents}$}{
				\uIf{$v_{pa}$ is $v_{origin}$}{
					append 0 to p\_list
				}
				\uElseIf{$v_{pa}$ has been visited already}{
					\If{$v_{pa} \in V_l$}{append $p(v_{pa})$ to p\_list}
					\Else{append 0 to p\_list}
				}
				\Else{append \texttt{RecursiveProbability(}$v_{pa}$, $v_{origin}$\texttt{)} to p\_list}
			}
			\If{$v \in V_a$}{
				\Return{\texttt{Conjunction(}p\_list\texttt{)}}
			}
			\If{$v \in V_o$}{
				\Return{\texttt{Disjunction(}p\_list\texttt{)}}
			}
		}
	}
\end{algorithm}
\end{minipage}

\begin{minipage}{\columnwidth}

\begin{algorithm}[H]
\removelatexerror
\texttt{Disjunction(}Probability List $P$\texttt{)}
	\tcc{Calculating the disjunctive probabilities for OR nodes}
	\KwIn{List of probabilities}
	\KwOut{Disjunction of the probabilities}
	p = $P$.pop(0)\;
	\uIf{length of $P$ is larger than 1}{
		recursive\_p = \texttt{Disjunction(}$P$\texttt{)}\;
		p = p + recursive\_p - recursive\_p $\times$ p\;
	}
	\uElseIf{length of $P$ is equal to 1}{
		\Return{p + $P$[0] - p$\times P$[0]}
	}
	\Else{\Return{p}}
	\Return{p}
\end{algorithm}
\end{minipage}

\begin{minipage}{\columnwidth}
\begin{algorithm}[H]
\removelatexerror
\texttt{Conjunction(}Probability List $P$\texttt{)}
	\tcc{Calculating the conjunctive probabilities for AND nodes}
	\KwIn{List of probabilities}
	\KwOut{Conjunction of the probabilities}
	p= $P$.pop(0)\;
	\uIf{length of $P$ is larger than 1}{
		recursive\_p = \texttt{Conjunction(}$P$\texttt{)}\;
		p = recursive\_p $\times$ p\;
	}
	\uElseIf{length of $P$ is equal to 1}{
		\Return{p $\times$ $P$[0]}
	}
	\Else{\Return{p}}
	\Return{p}
\caption{Propagating probabilities through the attack graph. The conjunctive and disjunctive functions correspond to calculating probabilities on OR and AND nodes, respectively.}
\label{alg:main}
\end{algorithm}
\end{minipage}
\vspace{-34pt}
\end{figure}


The input to the algorithm is the graph with local probabilities. The \texttt{pop(int)} function used in the \em{Disjunction} and \em{Conjunction} functions in Algorithm \ref{alg:main} is a list function that returns the item, in this case a probability, at the given list index then removes it from the list.

This algorithm achieves a calculation for node probabilities that makes sense given the context of network security without the normal requirement for removing edges from the graphs. This means that attack routes on graphs can be better understood while also improving the versatility of graphs as extra portions can be added and the new nodes can be calculated correctly as no edges have been removed.


\subsection{Complexity of the Algorithm}
The complexity of Algorithm~\ref{alg:main} can be calculated as $O(n\times \max_v |Pre(v)|)$ where $n$ is the number of nodes in the attack graph. In the worst case, when every added node is required to be present in the calculation of every other node, the complexity of the algorithm is $O(n^2)$. If the cyclicity of the graph is known, then the complexity becomes $O(n(cn + \max_v |Pre(v)|))$ where $0 \le c \leq 1$ and $c$ is the portion of nodes that are in at least one cycle. 


\subsection{Selection of Local Probabilities}
In order to infer the access probabilities for the nodes passed to the algorithm, an initial set of local probabilities must be provided. These were originally generated from a simplistic evaluation of the ease to exploit a vulnerability (with non leaf local probabilities set to 1 as discussed earlier in Section 3.2). In order to determine the ease of access, the CVSS vector \cite{First2015common,Mell2006common} for the vulnerability is collected from NIST's National Vulnerability Database (NVD). Currently the Access Complexity (CVSSv2) or the Attack Complexity (CVSSv3) is used to define the probability of transitioning to a state. This is on a scale of Low, Medium and High for version 2 and Low and High for version 3, with High meaning there is a great deal of skill or timing required to exploit the vulnerability and as such is associated with the lowest probability scoring. A demonstration of how these values could inform the local probabilities is shown in Table \ref{tab:scores}.

\begin{table}
    \caption{Complexity scores and their local probabilities.}
    \label{tab:scores}
    \centering
    \begin{tabular}{ccl}
        \toprule
	    Vector Score & CVSS Version & Local Probability \\  
        \midrule
	    Low/L & 2,3 & 0.71\\
	    Medium/M & 2 & 0.61\\
	    Unknown & - & 0.61\\
	    High/H & 2,3 & 0.35\\
        \bottomrule
    \end{tabular}
\end{table}

The local probabilities are taken from the contribution that the NVD gives to a vector score when calculating the whole CVSS score. While this is a useful approximation, it is very abstract and ignores a great deal of the information that can be gleaned from the information available about the vulnerabilities. A discussion of this can be found in Section \ref{sec:related}.

\section{Experimental Results}
\label{sec:experiments}

\subsection{Application to Simulated Networks}
 In order to test the practicality of the algorithm, it was implemented in Python, alongside a simulator that can generate attack graphs with cycles. This simulator builds a random attack graph with a specifiable quantity of cycles; it is given a percentage of cycles to artificially add, and ensures that the given percentage of OR nodes are involved in cycles (as this is where cycles originate, from the state of privileged access that allows potential future access to a vulnerability that has already been exploited). The graph is built out of nodes generated with a Leaf:AND:OR ratio of 50:35:15 in order to model the fact that approximately half a common attack graph comprises of configuration Leaf nodes, and there are fewer nodes representing states than there are representing attack steps (this is due to the fact that multiple different AND nodes can lead to the same OR node i.e. many actions can lead to the same state).

\begin{figure}
	\centering
	\includegraphics[clip,width=\linewidth]{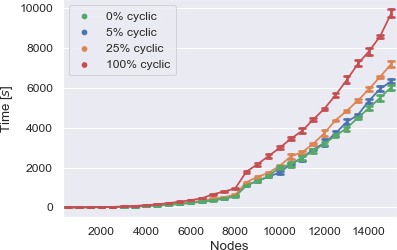}
	\caption{Time performance of Algorithm~\ref{alg:main}.}
	\label{fig:timeperf}
\end{figure}

Attack graphs were simulated in increasing sizes, from 500 to 15000 in step sizes of 500, and in four different groupings; `0\% cyclic' where there are no cycles in the graph, `5\% cyclic' where five percent of OR nodes are involved in cycles, `25\% cyclic', and `100\% cyclic' where every node, excluding Leaf nodes, is included in a cycle. Every situation was simulated five times, and the computation time required to complete exact inference using the algorithm was timed. These results are shown in Figure \ref{fig:timeperf}. Plotted are the average values for each amount of nodes, as well as range bars for the minimum and maximum values.

   As can be seen, the algorithm can generate probabilities for graph sizes of at least 15000 nodes in the worst possible case in a moderate time frame, around 155 minutes. Thus the algorithm is suitable for use on medium-large sized networks, especially if modeling techniques like consolidating similar work stations into singles nodes is used. By way of comparison, an enterprise with 300 machines, each with an average of 5 vulnerabilities, would create a graph of around 6000 nodes. In a worst case situation regarding cycles, the probabilities could be calculated in approximately 600 seconds.
\begin{figure}
	\centering
	\includegraphics[clip,width=\linewidth]{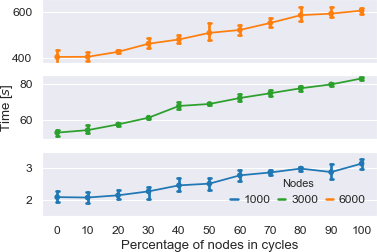}
	\caption{Impact of cyclicity on the computation time.}
	\label{fig:cycperf}
\end{figure}

The effect of increasing the percentage of cycles at fixed node quantities was also examined, with results displayed in Figure \ref{fig:cycperf}. Graphs were simulated with 1000, 3000 and 6000 nodes, with the percentage cycles being part of cycles is increased from 0 to 100 in steps of 10. The contribution of the cycles to the computation time increases with the quantity of cycles on the graph, up to an approximate 75\% increase in computation time in the worst case in the ranges of nodes that we experimented with. This can be seen in both Figures~\ref{fig:timeperf}-\ref{fig:cycperf} by comparing the results for 0\% cyclicity with 100\%. Including cycles seems like a somewhat expensive addition to graphs; however when any changes are to be modelled there will be no requirement to recompute the graph and as such the upfront cost will mean that only small portions of the graph will have to be computed after changes are made as the structure of the graph will be correct. In other words the upfront increase in time significantly reduces the cost of future adaptations and analyses of the graph.

\subsection{Application to Realistic Networks}
To confirm the applicabality of the results to realistic scenarios, we have implemented the algorithm on a collection of realistic networks. These are networks designed to replicate common real-life networking implementations for testing and modelling purposes. They are set up as a combination of virtual and real assets that function as an enterprise network, with different numbers of hosts and different vulnerabilities on each machine. These network scenarios return the same scan results that one would expect from running a vulnerability scan on the real network equivalent of the scenario.

We have considered three networks with respectively $10$, $15$, and $15$ hosts; the first network comprises of a Linux database and workstation, an iOS device, a peripheral device that uses SMB and a Windows server, with all the remaining hosts being Windows workstations. The workstations have various commonly used applications installed (Chrome, iTunes, JDK) and are in various patching states. The other networks have similar enterprise topologies, see Appendix~\ref{app:real} for more details. Each network is scanned using a modified version of the OpenVAS vulnerability scanner \cite{openvas}, and then an attack graph is generated for each using MulVAL \cite{Ou2005:MLN:1251398.1251406}. The number of nodes in the attack graphs generated from these networks is $1053$, $2234$ and $2341$; all have naturally occurring cycles. This gives us attack graphs that are roughly $5\%$ cyclic. The mean runtime of our algorithm is $3$, $11$, and $12$ seconds, respectively. This runtime confirms the quantities reported in Figure~\ref{fig:timeperf}: the simulations for attack graphs with $5\%$ cyclicity and $1000$ and $2000$ nodes have mean runtimes of $2$ and $18$ seconds respectively when run on the same machine used for the realistic networks.

\subsection{Evaluation on Examples from Literature}
To demonstrate the validity of our method we have applied Algorithm \ref{alg:main} on two common attack graphs from the literature;
an acyclic graph generated from a scenario presented by Wang et al.~\cite{wang2008attack}, and the cyclic example in Figure \ref{egbag} that was created by Ou et al.~\cite{Ou2006:SAA:1180405.1180446} and had probabilities calculated by Homer et al. by creating a larger equivalent graph that is acyclic \cite{homer2013aggregating}.
The full demonstration of these comparisons can be seen in Appendix~\ref{app:exp}.
%
Once the acyclic example was converted from a plain BAG into an AND/OR BAG, our algorithm was run on both examples. The results were identical to the expected results in the other papers, demonstrating that this algorithm gives correct results for common network scenarios and generalises them to the larger class of cyclic BAGs.

\subsection{Comparison with Exact Methods}
\begin{figure}
	\centering
	\includegraphics[clip,width=\linewidth]{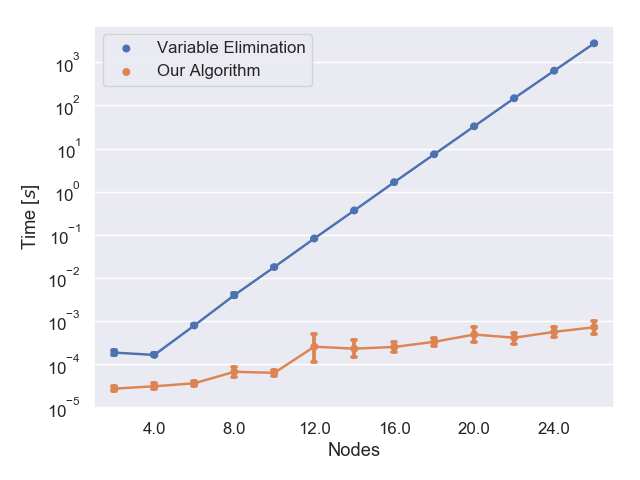}
	\caption{Comparing computational time of our algorithm with Variable Elimination on acyclic graphs.}
	\label{fig:veperf}
\end{figure}
In order to verify and compare our results on acyclic graphs, we implemented the Variable Elimination algorithm of Section~\ref{sec:VE} on the simulated graphs that are acyclic. Due to the limitations of Variable Elimination, we implemented it on a series of small simulated graphs from 2 to 26 nodes, and compared the results with our algorithm. The average error of the quantities computed by our algorithm is $\pm0.011$.

As can be seen in Figure~\ref{fig:veperf}, Variable Elimination despite being exact scales very poorly with one run on a $24$ node graph taking approximately the same amount of time as a $11500$ node run with our algorithm. Due to this poor scaling, the fact that Variable Elimination cannot run on cyclic graphs, and the low average error from our algorithm, our approach is a much better choice for all practical applications.
\section{Related Work}
\label{sec:related}
Network vulnerability analysis using graph based approaches has been widely covered in the literature. A large body of work exists using attack graphs which propose approaches and tools for improved generation \cite{Ingols2009,Ou2006:SAA:1180405.1180446,Sheyner2002,Sheyner2004scenario,Swiler2001}, visualization \cite{Homer2008,Lee2018}, and analysis \cite{Dewri2007:OSH:1315245.1315272,Ammann2002:SGN:586110.586140,Roschke2011}, or provide summaries of existing techniques \cite{Barik2016,Kaynar2016,Kordy2014,Yi2013}.   Of the many types of attack graphs that exist, we have used dependency attack graphs, which facilitate understanding of the importance of individual vulnerabilities \cite{Sawilla2007googling}.  For a complete overview of the different techniques as well as a taxonomy for their formalisms we refer to Kordy et al.~\cite{Kordy2014}.

 Bayesian network based network analysis has also been studied (e.g. \cite{Frigault2008:MNS:1456362.1456368,liu2005network,Poolsappasit2012,sembiring2015network,Xie2010}) as well as their generation. In particular, Choi et al. propose a method for deleting network edges from a Bayesian network in order to generate models through approximate inference \cite{Choi2014}. In \cite{Munoz-Gonzalez2016}, Mu\~{n}oz-Gonz\'{a}lez et al. apply an approximate inference approach called loopy belief propagation to Bayesian attack graphs and shows its performance compares favourably to the widely used exact inference technique, Junction Tree (JT). In \cite{Munoz-Gonzalez2017} the authors explore an exact inference method using network clustering which they show enables the JT algorithm to become tractable and scale linearly with the number of nodes.

In order to improve the accuracy of these graphs, insightful local probabilities should be generated to maximise the information available to the Bayesian network. Doynikova and Kotenko demonstrate in \cite{Doynikova2017} a more complex method for achieving more accurate results from CVSS data than simply using the complexity score, while also modelling attacks that do not rely on vulnerability exploitation through the use of the CAPEC list (Common Attack Pattern Enumeration and Classification), a taxonomy of different attack patterns described using the MITRE schema \cite{Barnum2008common}. Cheng et al. model dependency relationships of the base metrics in the vectors and attempt to combine them in such a way that a user can weigh specific aspects for their local probability assignment \cite{Cheng2017}.

A number of Markovian approaches have been taken to generate Bayesian attack graphs and facilitate vulnerability analysis and the design of optimal defence strategies. Jha et al. use a model checker to automatically generate attack graphs annotated with probabilities and analyse their vulnerabilities using Markov Decision Process (MDP) algorithms \cite{Jha2002}. Mace et al. used a similar approach to find the optimal data collection strategies for accurate Bayesian attack graph input parameters (e.g. conditional probability tables) \cite{mace2017}. In \cite{PietreCambacedes2010} Pi\`{e}tre-Cambac\'{e}d\`{e}s et al. model attack trees as Boolean logic Driven Markov Processes (BDMP), suggesting they are dynamic and inherit readability and appropriation of attack trees but with mathematical properties reducing combinatorial problems and processing. Continuous Time Markov chains have been applied by Jhawar et al. to the Bayesian attack graph approach in order to analyse attack defence graphs, that is, attack graphs which define the modelling of defences in their specification. Wang et al. in \cite{wang2008attack}, estimate attack states and define a cost-benefit heuristic to automatically infer optimal defences for attack graphs integrated with Hidden Markov Models while Miehling et al. \cite{Miehling2015:ODP:2808475.2808482} and Zhisheng et al. \cite{hu2017online} assume the defender can only partially observe the attacker's capabilities at any given time, thereby modelling Bayesian Attack Graphs as partially observable Markov decision processes (POMDPs). In this sense a resulting defence strategy is both reactive and anticipatory. 

Dealing with cycles in Bayesian attack graphs has also been tackled (e.g.\cite{Peng20105544924,Jiang2018}). K\l opotek et al. who use Markov Chains, suggesting the state of a variable is not influenced by itself but rather the future state is influenced by the past one \cite{klopotek2006}. Doynikova and Kotenko consider the processing of three simple types of cycle in Bayesian attack graphs \cite{Doynikova2017}. Two of the three types contain cycles between nodes at the same level in the graph structure. These are either removed once the probabilities to reach each node for the first time have been calculated, or enumerated as separate paths. The third type contains cycles between nodes at different levels of the graph structure. In this case, the cycle is simply removed under the backtracking assumption, that is an attacker does not come back to a node already exploited. In \cite{Aguessy2016}, Aguessy et al. present a Bayesian network-based extension to attack graphs, called a Bayesian Attack Model (BAM), which is capable of handling cycles by breaking them. The authors argue that using the backtracking assumption to break cycles suppresses possible attacker actions which cannot be known a priori. To keep all possible paths, the only way to break cycles is to enumerate all paths starting from every possible attack source. In other words, unfolding the cyclic graph structure to an equivalent acyclic graph structure such that each node appears exactly once in each path. This process causes a combinatorial explosion in the number of nodes whilst the inference algorithm is shown to remains efficient only for networks of up to 70 hosts. Homer et al. suggest their approach correctly handles both cycles and shared dependencies in attack graphs, that is the probabilities along multiple paths leading to a node are dependent on each other \cite{homer2013aggregating}. The authors suggest enumerating all paths is unnecessary if data flow analysis is applied to the cyclic nodes enabling the same probabilities to be evaluated as on the unfolded graph. The data flow analysis process uses dynamic programming and other optimizations to avoid increasing computational complexity. The algorithm is limited by the number of nodes and paths within cycles which must be considered when calculating probability values and which can cause evaluation time to be exponential in the worst case. The evaluation is based on the number of nodes and vulnerabilities per node and does not consider how the number of cycles impacts computation. Wang et al. made a number of crucial observations about cyclic attack graphs and proposed a customized probabilistic reasoning method that can handle cycles in the calculation \cite{wang2008attack}. However, when combining probabilities from multiple attack paths, the method uses a formula that assumes the multiple probabilities are independent. Such dependency needs to be accounted for to prevent a distortion of results.

\section{Conclusion}
\label{sec:conclusion}

In this paper we have created and demonstrated a systematic approach to analyse Bayesian attack graphs, including those with cycles.  Since cycles naturally arise in BAGs that are generated from scanning software (e.g., using MulVAL), it is imperative to establish practical approaches to handle cyclic BAGs.  We presented a formal treatment of the problem domain and introduced a solution algorithm that can be applied to any BAG, cyclic or acyclic.  This results in a method by which Bayesian attack graphs can be automatically evaluated with respect to what states are available to an attacker and how easily they are reached. Our solution deals directly with the cycles and integrate cycle resolution with the computation of state probabilities without the need for identifying or differentiating the cycle types.
Our approach \emph{does not} alter the attack graph by removing edges to make it acyclic. Instead, we preserve all the information in the graph, and no potential attack routes are lost when new data is added to the graph.

Our computational approach is currently restricted to single state probabilities and cannot compute joint probabilities for multiple nodes. A solution that allows the computation of joint probabilities is a next step to further advance the presented approach to cyclic BAGs. Future work will also involve extending the local probability assignment to have a more meaningful value; a temporal metric can be included given that the longer a vulnerability is known about the more likely an exploit has been published for it, increasing the ease of an attack. Scaleable solution algorithms then need to be identified to automate the analysis of these graphs to prioritise fixing of vulnerabilities and identifying most vulnerable network hosts, with regard to their criticality to an enterprise. Finally, although the proposed algorithm can handle large models, for yet larger networks approximate solutions could be considered.

\bibliographystyle{ieeetr}
\bibliography{./references}

\appendices

\section{Attack Graphs with Cycles}
\label{sec:cycles}
\begin{figure*}
	\centering
	\begin{minipage}{.2\textwidth}
		\centering
		\includegraphics[width=0.775\linewidth]{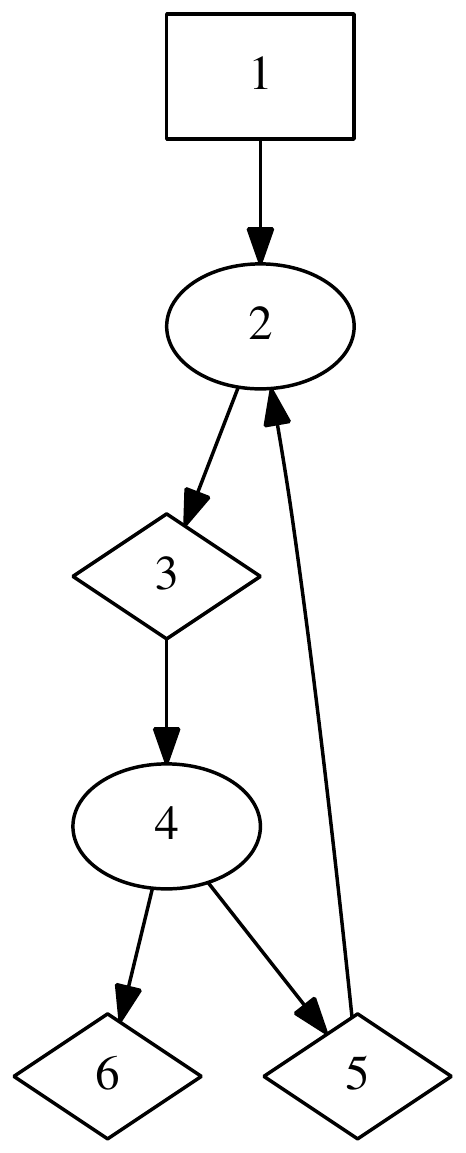}
		\caption{Cycle of Type 1.}
		\label{fig:loopone}
	\end{minipage}
	\quad\quad\quad
	\begin{minipage}{.2\textwidth}
		\centering
		\includegraphics[width=0.6375\linewidth]{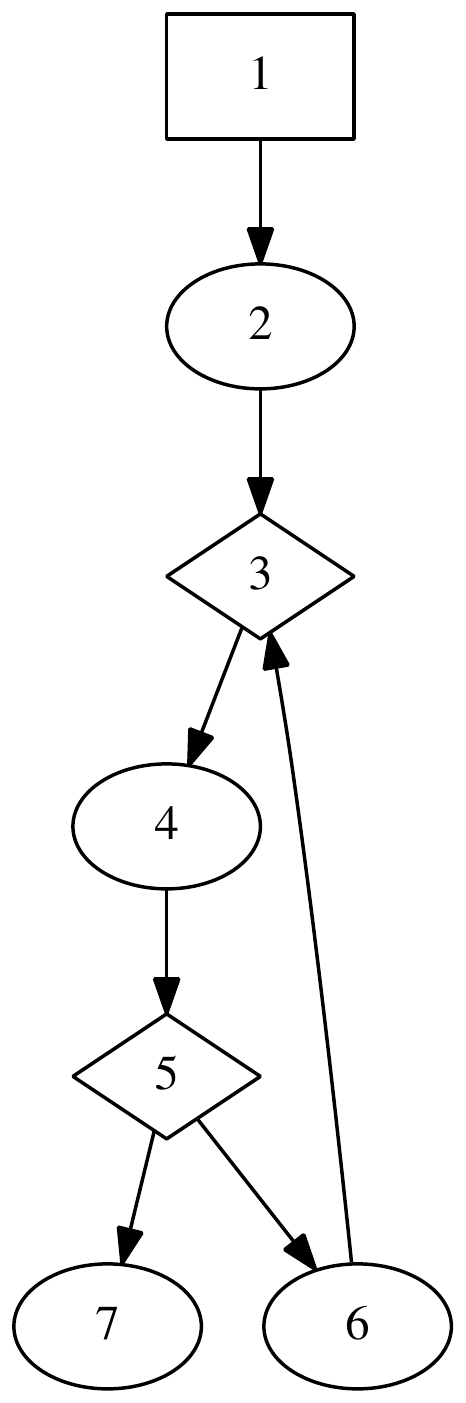}
		\caption{Cycle of Type 2.}
		\label{fig:looptwo}
	\end{minipage}
	\quad\quad\quad
	\begin{minipage}{.24\textwidth}
		\centering
		\includegraphics[width=0.749\linewidth]{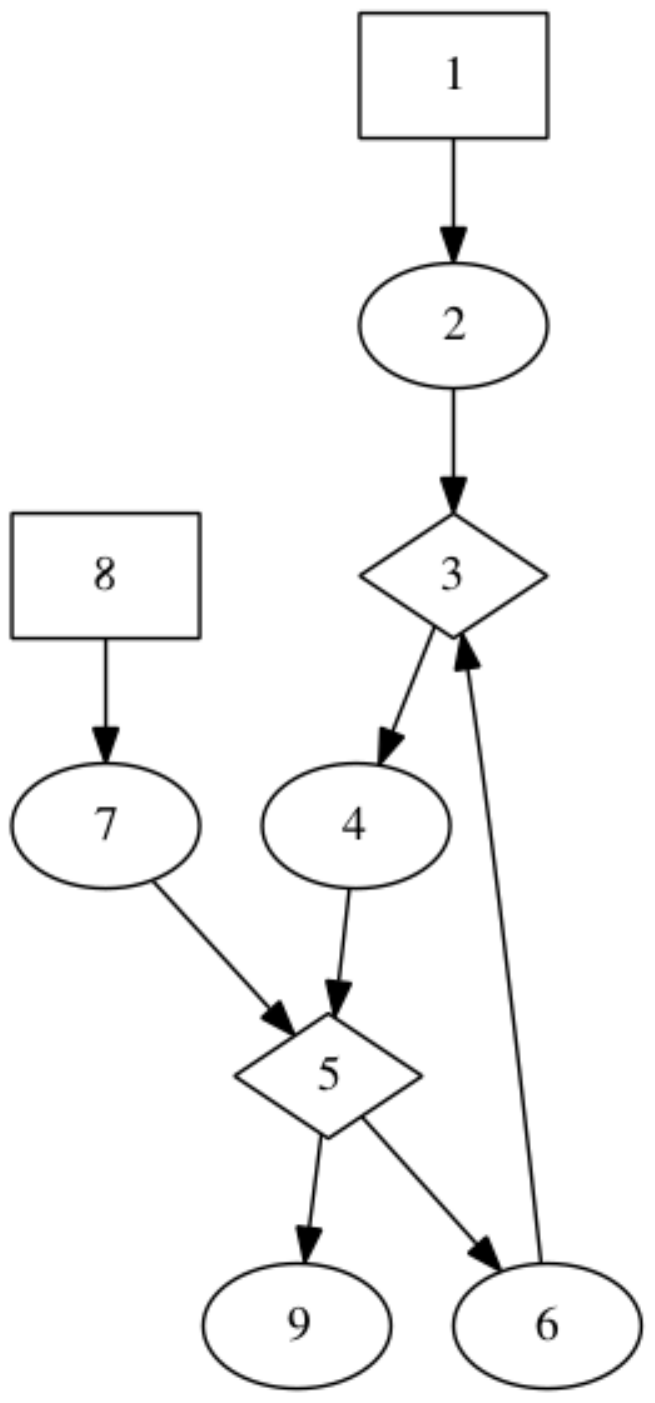}
		\caption{Cycle of Type 3.}
		\label{fig:loopthree}
	\end{minipage}
\end{figure*}
As we discussed in Section~\ref{sec:motivation}, most of the literature on BAG probability computations have focused on acyclic attack graphs. This constraint ensures the probabilities on the nodes become the chance that an attacker reaches the node at all. In other words, paths that are enabled by access to a node should not increase the probability that that same node is reached. We presented a running example and showed that cycles can occur in a number of situations. Thus, this property should be extended for each node on the pathway calculations.

In the following, we first discuss types of cycles mentioned in the literature and the methods currently used to deal with these cycles. Next we put these types of cycles into the perspective of our computational algorithm and show how they can be interpreted over the associated combinational circuit. 
We emphasise that our solution does not modify the graph in any way while being able to run on graphs containing cycles. Our solution deals directly with the cycles and integrate cycle resolution with the computation of state probabilities without the need for identifying or differentiating the cycle types.

\subsection{Handling Cycles}





There are three main types of cycles in an attack graph \cite{Frigault2017,Doynikova2017}, and Figures \ref{fig:loopone} to \ref{fig:loopthree} show what they look like based on Definition~\ref{def:BAG2}. The first cycle type, Figure \ref{fig:loopone}, can be demonstrated as removable. This is because node 2 has two prerequisites and one of them, node 5, is only fulfilled given that node 2 has been accessed. As such node 2 can never be true and the cycle will never occur and does not aid our understanding or modelling of the graph.

Figure~\ref{fig:looptwo} shows the second type of cycle, one that cannot be trivially removed like the first. Node 3 can be fulfilled by either 2 or 6 meaning that node 7 is reachable. Since node 6 can only be accessed after node 3 has been accessed, it should not contribute to the likelihood of reaching nodes 3, 4, 5 or 7. In essence the edge from node 6 to 3 could be removed from the graph, and this is the necessary step in current acyclic BAG techniques.

An important addition to this discussion is that while edge removal does work for the second type of cycle, it is perhaps not the preferred solution. Firstly removing the edge can make the graph less understandable from an engineering perspective, as logically if reaching a state allows  access to another element that was a prerequisite to a previous state then that route is possible. Also, more importantly, removing an edge, the edge from node 6 to node 3 in our example, removes information that could be important in future. If it was desired that a new vulnerability be added to the graph, for example in such a way that it would transform Figure \ref{fig:looptwo} into Figure \ref{fig:loopthree}, then a legitimate route would now be missing from the graph and from any new calculations. As such, preserving the structure of the graph is important for both future analysis and understanding.

The third type of cycle, Figure \ref{fig:loopthree}, shows a cycle that cannot be ignored or fixed through edge removal. It is the same in structure as the Type 2 cycle but a node in the cycle has an extra way of being accessed, meaning there are now multiple routes to reach node 3. This type of cycle should be dealt with by imagining probabilities are populations of attackers and as such the quantities on the nodes should represent unique attacker numbers ensuring we do not double count attackers moving through the graph. Practically for this simplistic example this will mean the probability at node 6 is the disjunction of attackers coming from node 7 and the attackers reaching node 4 for the first time, i.e. the initial population at node 1 minus any attackers lost moving through nodes 2 and 3.

\subsection{Cycles in Combinational Circuits}
Using the combinational circuit paradigm introduced in section 4, we can formally describe the different types of cycle.
The finiteness of $k^\ast$ mentioned in Theorem~\ref{thm:finite_k} enables us to unfold the logic circuit $k^\ast$ number of times.
Let us denote 
$$pa(v_i) = \{v_{i1},v_{i2},\ldots,v_{i,m_i}\}.$$
The unfolding is done sequentially by replacing each $v_{ij}(k)$ in the right-hand side of \eqref{eq:dynamical_logic} with function $g_z(pa(v_z)(k-1),v_z')$ where $v_z$ is the node associated with $v_{ij}$. Starting this process from $k^\ast$ and repeating it $k^\ast$ times gives us a full circuit as
\begin{equation}
\label{eq:dynamical_logic2}
\left\{
    \begin{array}{l}
       v_1(k^\ast) = f_1(v'_1,v'_2,\ldots, v'_n)\\
        v_2(k^\ast) = f_2(v'_1,v'_2,\ldots, v'_n)\\
        \quad \vdots\\
        v_n(k^\ast) = f_n(v'_1,v'_2,\ldots, v'_n),
    \end{array}
    \right.
\end{equation}
where $f_i$'s are associated to the unfolded circuit with a directed graph that does not have any cycles.
Unfortunately, the procedure of unfolding and probability computation over \eqref{eq:dynamical_logic2} is computationally intense but it is very helpful in giving an automatic characterisation of cycle types in BAGs discussed in the literature \cite{Frigault2017,Doynikova2017}.

Cycles of Type 1 are seen when the steady-state value of a node is zero: $v_i(k^\ast)=0$ for some $i$ and for any instantiation of $\{v'_1,v'_2,\ldots, v'_n\}$. This means that the node cannot be reached and can be safely eliminated from the analysis. Any incoming edges or outgoing edges to this node can also be eliminated. If this elimination results in breaking a specific cycle, that cycle is of Type~1.

Cycles of Type~2 need more elaboration and are defined with respect to nodes that $v_i(k^\ast) = 1$.
Let $k^\ast_i$ be the earliest time that the value of node $v_i$ becomes one:
\begin{equation}
    k^\ast_i = min_k\{k, v_i(k) = 1\}\text{ for } i \text{ with } v_i(k^\ast) = 1.
\end{equation}
It is obvious that $k^\ast_i$ depends on the instantiation of $\{v'_1,v'_2,\ldots, v'_n\}$ and is upper bounded by $k^\ast$. 
The computation of $\text{prob}(v_i(k^\ast) = 1)$ requires unfolding \eqref{eq:dynamical_logic2} for $k^\ast_i$ times. Nodes with the property that $v_j(k^\ast_i) = 0$ can safely be removed together with their outgoing and incoming edges. These are the nodes that do not have any influence on node $v_i$. If such elimination results in breaking a cycle, that cycle is of Type~2. Note that the elimination is valid when we only need to compute access probabilities of $v_i$ (the definition is with respect to a particular node).

The previous two types of cycles require properties that should hold for any instantiation of $\{v'_1,v'_2,\ldots, v'_n\}$. Cycles of Type~3 are the ones that do not fit in the definition of cycles of Type~2. Formally, for a given node $v_i$, a cycle is of Type~3 if there exists a node $v_j$ on the cycle and an instantiation of $\{v'_1,v'_2,\ldots, v'_n\}$ such that $v_j(k_i^\ast-1) = 1$. This means node $v_j$ on the cycle can influence the access probability of node $v_i$, thus cannot be removed.

As discussed above, cycles of Type~3 cannot be removed and requires a particular attention when performing the probability computations. In the next subsection, we demonstrate the computation on the running example and provide the full algorithm in Section~\ref{sec:calculation}.

\begin{figure*}
	\centering
	\includegraphics[width=0.6\linewidth]{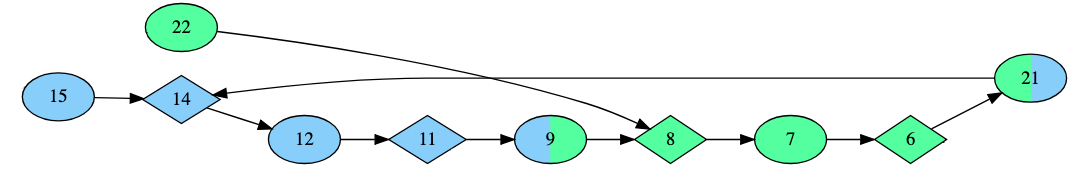}
	\caption{The cycle of the BAG presented in Figure \ref{egbag}.}
	\label{fig:cycleexcerpt}
\end{figure*}
\begin{figure*}
\centering
	\includegraphics[width=0.75\linewidth]{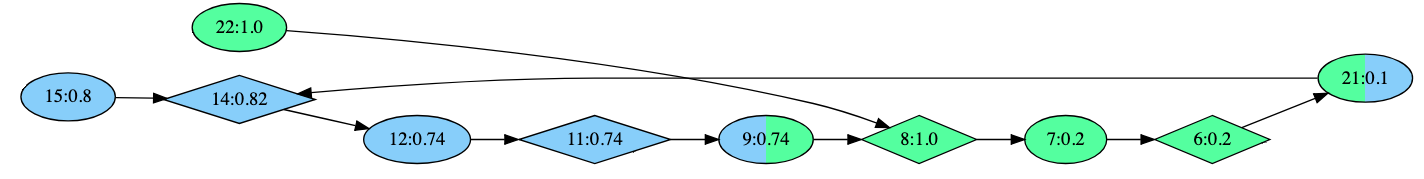}
	\caption{The cycle of the BAG presented in Figure \ref{egbag} with computed access probabilities.}
	\label{fig:cycleexcerptcalc}
\end{figure*}
\subsection{Calculating Probabilities: Running Example}

Here we discuss how a cycle's probabilities should be calculated, as is implemented in the algorithm in the following section.
Figure \ref{fig:cycleexcerpt} is an excerpt from the larger attack graph of the running example presented in Section~\ref{sec:running_example}. It is a simple example of how a Type 3 cycle (see Figure \ref{fig:loopthree}) can exist in a real attack graph. This excerpt is a trivial demonstration of the simplest occurrences of cycles in real attack graphs. The cycle occurs because of the multiple routes that can be taken to reach node 14, where an Internet Explorer vulnerability on the Workstations can be exploited using an HTML document. A user may visit a malicious website, represented by the route from node 15 to 14, or alternatively the Webserver could be targeted first, and used to provide the HTML document once it has been compromised, shown in the route through nodes 8,7,6,21 and then 14.

The cycle is further complicated by the fact that the Webserver can be accessed directly without going through the Workstations, in the route from node 22 to 8. Without the edge $e_{22,8}$, the edge causing the cycle ($e_{21,14}$) could be safely trimmed as the probability of the attacker reaching node 14 does not increase due to node 21 as node 14 is a prerequisite for node 21 to be reached.  Node 22 entering the cycle part way through, however, will increase the probability of reaching node 14 at some point in an attack as node 15 being accessed is no longer a requirement.

The result of calculating the probability of reaching each node, performed by disregarding nodes that have already contributed, can be seen in Figure \ref{fig:cycleexcerptcalc}. In order to calculate probabilities within the cycle, all the parent nodes are collected exhaustively. Their contribution to the probability of the node in question is then performed according to the relationships defined by the graph, as in definition \ref{def:score2}, with the caveat that any node in the parent set may only contribute once to the calculation. In this way, calculating the probability of node 12 on Figure \ref{fig:cycleexcerpt} will involve the likelihood of any attacker reaching node 14 from node 15, and also the likelihood of an attacker reaching node 14 from node 21, but with the removal of node 15's contribution to the probability of reaching node 21 as node 15 has already been included. In this way each nodes probability can be calculated without the removal of any edge, as a node causing a loop in one place may contribute to probabilities elsewhere on the graph and as such should only be disregarded in specific calculations where it's effects have already been calculated. The ability for a node to be present in multiple paths but contribute differing amounts demonstrates the idea that a recursive algorithm that identifies each node's contribution to a path would be a correct solution to this problem, preventing any node from contributing multiple times to the same path.

\section{Full Example}
\label{app:eg}
\begin{figure*}[ht]
    \centering
	\includegraphics[width=\linewidth]{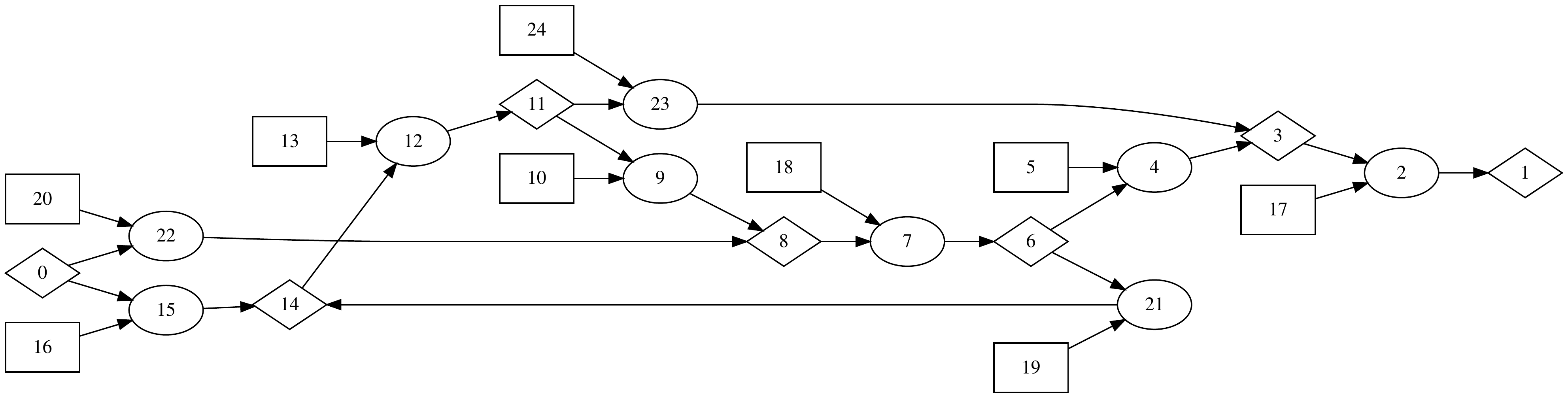}
	\caption{The BAG of the running example including leaf nodes.}
	\label{fullbag}
\end{figure*}
The complete attack graph for the running example scenario can be seen in Figure \ref{fullbag}, with the labels for the nodes written out below. An important note is the reason the cycle exists: the state at node 14, whereby a user on one of the Workstation machines access a malicious website, can be reached via two means. Firstly the user can simply visit a malicious website allowing the attacker to exploit CVE-2009-1918 that is in Internet Explorer on the Workstation. Alternatively, an attacker that has achieved the ability to execute code on the Web Server (node 6) can serve the user of a Workstation machine an HTML document that exploits CVE-2009-1918 and thus also achieves the state on node 14.

This cycle is made more complex due to the ability to access the Webserver without passing through the Workstations originally (visiting nodes 22 and 8). Because of this, reaching node 14 via node 21 will not always mean that the attacker is travelling backwards, and as such the monotonicity principle does not apply and the probability of reaching node 14 becomes more challenging to calculate.

The vulnerabilities in this scenario are as follow:
\begin{itemize}
    \item CVE-2009-1918\footnote{https://nvd.nist.gov/vuln/detail/CVE-2009-1918} on the Workstations - Internet Explorer vulnerability that allows an attacker to execute arbitrary code on the machine after the user accesses a website with purposely malformed elements that trigger memory corruption
    \item CVE-2006-3747\footnote{https://nvd.nist.gov/vuln/detail/CVE-2006-3747} on the Webserver - Apache vulnerability that can be exploited to execute arbitrary code using crafted URLs and requires network access to exploit
    \item CVE-2009-2446\footnote{https://nvd.nist.gov/vuln/detail/CVE-2009-2446} on the Database Server - MySQL vulnerability where an authenticated user can cause a denial of service and possibly execute arbitrary code
\end{itemize}

\lstset{
  basicstyle=\ttfamily,
  columns=fullflexible,
  keepspaces=true,
}
\begin{lstlisting}[caption=MulVAL labels for Figure \ref{fullbag}, label= labellistin, basicstyle=\small\ttfamily]
0, "attackerLocated(internet)"
1, "execCode(dbServer,root)"
2, "RULE 2 (remote exploit of a server 
program)"
3, "netAccess(dbServer,tcp,'3306')"
4, "RULE 5 (multi-hop access)"
5, "hacl(webServer,dbServer,
tcp,'3306')"
6, "execCode(webServer,apache)"
7, "RULE 2"
8, "netAccess(webServer,tcp,'80')"
9, "RULE 5"
10, "hacl(workStation,webServer,tcp,'80'"
11, "execCode(workStation,userAccount)"
12, "RULE 2"
13, "vulExists(workStation,'CVE-2009-1918',
IE,remoteExploit,privEscalation)"
14, "accessMaliciousInput(workStation,
user, IE)"
15, "malicious website"
16, "visit of malicious website"
17, "vulExists(dbServer,'CVE-2009-2446',
mySQL,remoteExploit,privEscalation)"
18, "vulExists(webServer,'CVE-2006-3747',
apache,remoteExploit,privEscalation)"
19, "visit of compromised website"
20, "hacl(internet, webServer, tcp, '80')"
21, "compromise of website"
22, "RULE 6 (direct network access"
23, "RULE 5"
24, "hacl(workStation,dbServer,tcp,'3306')
\end{lstlisting}

\section{Results of Algorithm on Common Examples}
\label{app:exp}
\subsection{Acyclic Example}
\begin{figure}[ht]
	\centering
	\includegraphics[clip,width=0.7\linewidth]{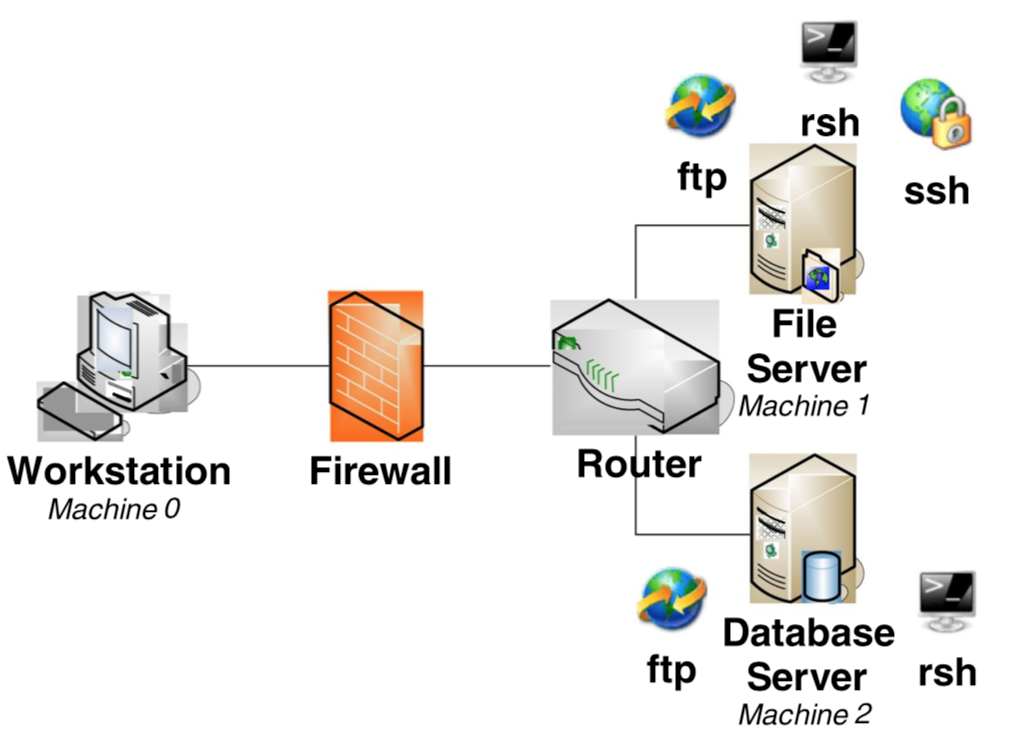}
	\caption{Example of a network taken from \cite{wang2008attack}.}
	\label{fig:ageg1}
\end{figure}

\begin{figure}
	\centering
	\includegraphics[clip,width=0.7\linewidth]{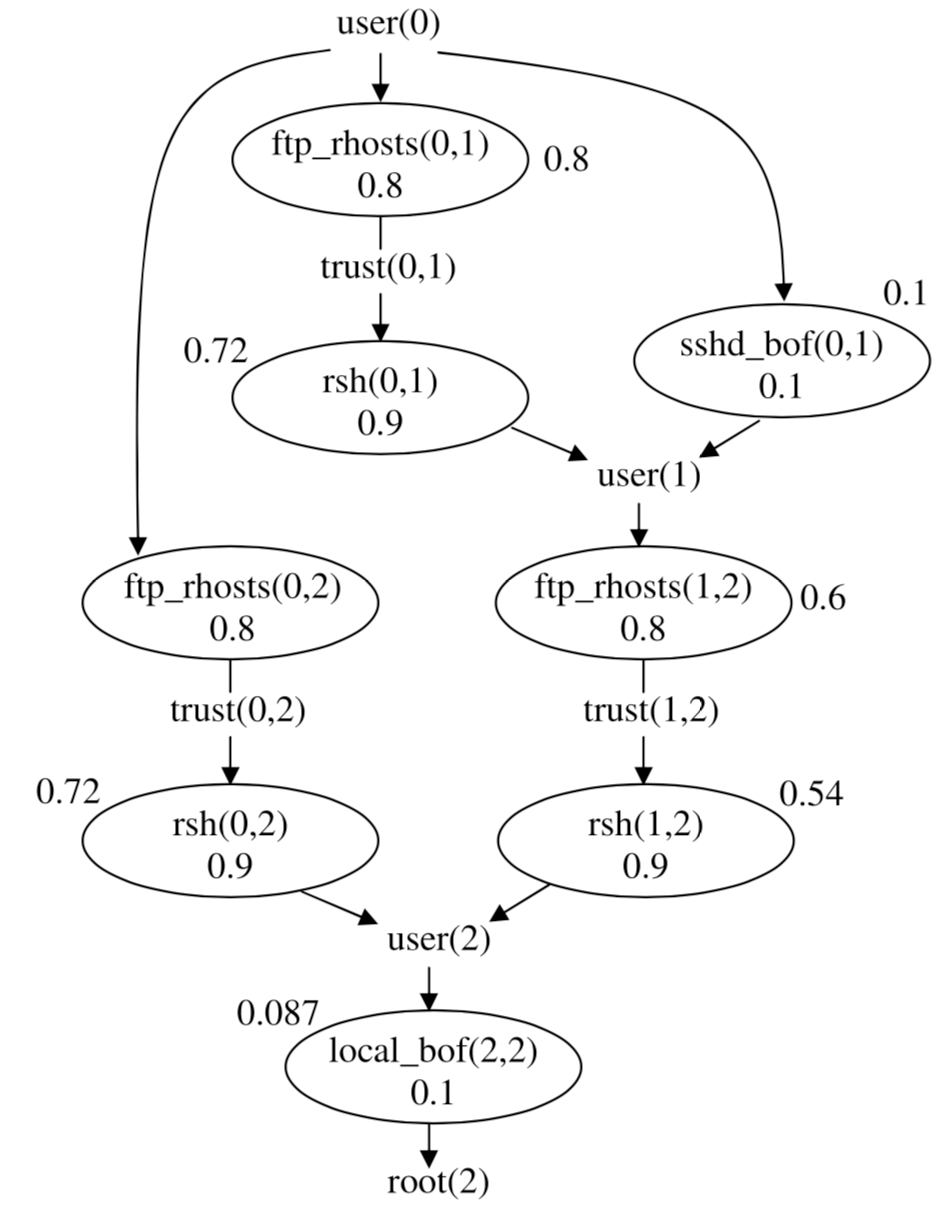}
	\caption{BAG of the network of Figure~\ref{fig:ageg1} which is acyclic \cite{wang2008attack}.}
	\label{fig:ageg2}
\end{figure}

The network scenario, in Figure \ref{fig:ageg1}, is moving from a Workstation to root access on a Database server through a Firewall and possibly via a File server depending on the attack path. There are three services running on Machine 1, the file server, and two services running on Machine 2, the target Database server. An attack graph has already been generated for this scenario but using a different schema, shown in Figure \ref{fig:ageg2}. Here it can be seen that there are three possible paths for achieving the goal of root access to Machine 2; the attacker can either edit the trusted host list on Machine 2 to gain enough access to the host in order to run the buffer overflow attack, or the attacker can attempt to reach the same privilege by changing the relationship between Machine 1 and Machine 2, either by initially editing Machine 1's trusted host list or by attempting a buffer overflow attack against Machine 1.

\begin{figure}
	\centering
	\includegraphics[clip,width=0.95\linewidth]{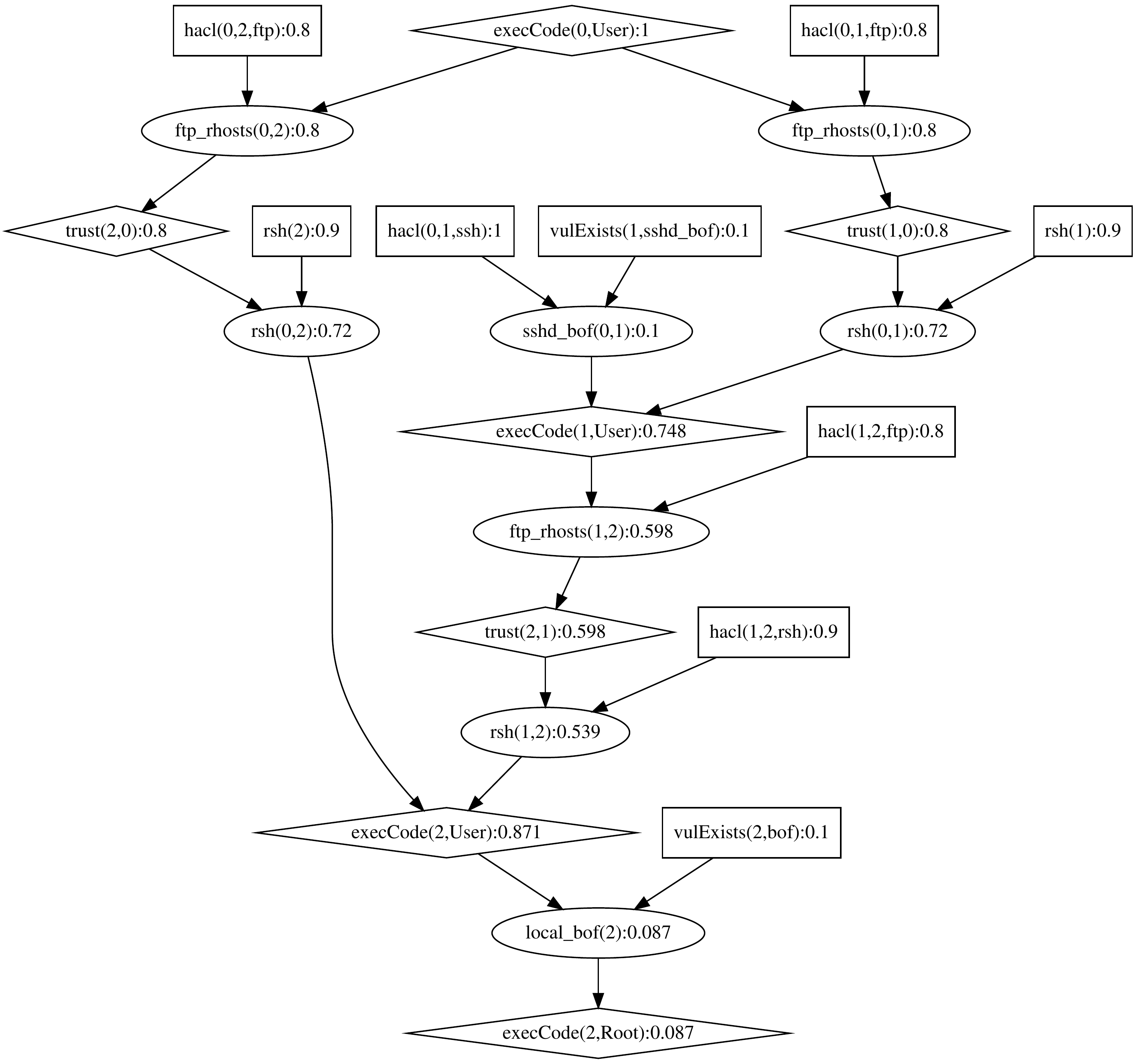}
	\caption{Converted graph from Figure \ref{fig:ageg2} using the equivalence shown in Section~\ref{sec:equivalence} }
	\label{fig:cbag}
\end{figure}

First, an AND/OR BAG was generated from the plain BAG in Figure~\ref{fig:ageg2} using the principles discussed in the paper. This involves moving all less than one local probabilities to the leaf nodes allowing easy logical calculation of the marginal probabilities at each node. The algorithm was then run on the new attack graph with the same probabilities associated with the vulnerabilities as allocated in the original example to generate a new Bayesian attack graph. This new graph, Figure \ref{fig:cbag}, shows the same probabilities for each part of the graph. This demonstrates that the algorithm is correct for this common acyclic graph example.

\subsection{Probability computations on the running example}
\begin{figure}
	\centering
	\includegraphics[clip,scale=0.5]{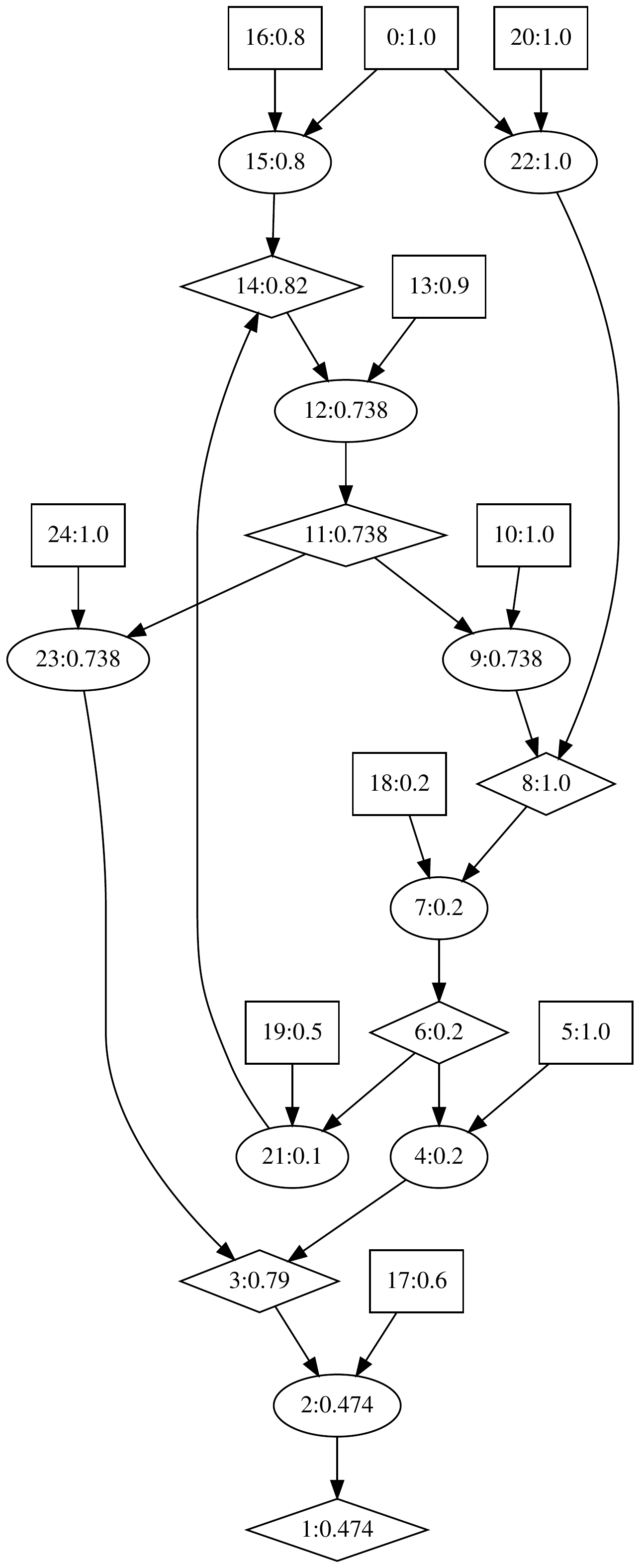}
	\caption{Results of Algorithm~\ref{alg:main} applied to the cyclic running example.}
	\label{fig:cycbageg}
\end{figure}

The probabilities are calculated and shown in Figure~\ref{fig:cycbageg} by applying Algorithm \ref{alg:main} to the running example presented in Section~\ref{sec:running_example} and described in appendix \ref{app:exp}. This graph has all the nodes displayed, including the leaf nodes that were trimmed for clarity through the rest of the paper.

\section{Realistic Networks}
\label{app:real}
\subsection{1053 Nodes}
This network is a simple small enterprise setup, with several workstations, some servers, and a collection of peripheral devices. The full host inventory can be seen in table \ref{tab:real1}.
\begin{table}[!htp]
    \caption{Hosts and Software for the 1053 Node Realistic Network.}
    \label{tab:real1}
    \centering
    \begin{tabular}{lcl}
        \toprule
	    Type & Amount & Software \\  
        \midrule
	    Windows Workstation & 4 & Internet Explorer, JDK, iTunes,\\
	    & & Office\\
	    Windows Server 2008 & 1 & -\\
	    SMB Device & 1 & -\\
	    Linux Machine & 2 & Pidgin, Chrome, Firefox, Samba\\
	    Linux Database Server & 1 & -\\
	    iOS Machine & 1 & Apple TV\\
        \bottomrule
    \end{tabular}
\end{table}
\subsection{2234 Nodes}
This network is a more complex enterprise example, and includes a server running TWiki that all the workstations can access for collaboration. The full host inventory can be seen in table \ref{tab:real2}.
\begin{table}[!htp]
    \caption{Hosts and Software for the 2234 Node Realistic Network.}
    \label{tab:real2}
    \begin{tabular}{lcl}
        \toprule
	    Type & Amount & Software \\  
        \midrule
	    Windows Workstation & 4 & Internet Explorer, JDK, Office,\\
	    & & DirectX, Edge\\
	    Windows Workstation & 3 & LiveMeeting, Edge\\
	    Windows Server 2008 & 1 & -\\
	    SMB Device & 2 & -\\
	    Ubuntu Machine & 2 & Pidgin, Chrome, Firefox, Apport,\\
	    & & Python, Jasper, OpenSSL,\\
	    & & Libxml2, Poppler\\
	    Linux Database Server & 1 & -\\
	    TWiki Web Server & 1 & TWiki, PCRE, PHP, Samba\\
	    Remote Login Machine & 1 & OpenSSH\\
        \bottomrule
    \end{tabular}
\end{table}
\subsection{2341 Nodes}
This network has similar hosts to the 2234 Node example, but introduces an outdated Windows XP machine to the network, along with a machine for web development and a Red Hat MRG machine. The full host inventory can be seen in table \ref{tab:real3}.
\begin{table}[!htp]
    \caption{Hosts and Software for the 2341 Node Realistic Network.}
    \label{tab:real3}
    \centering
    \begin{tabular}{lcl}
        \toprule
	    Type & Amount & Software \\  
        \midrule
        Old Windows Machine & 1 & Windows XP, Flash Player, \\
        & & JavaFX, Adobe Air, Wireshark\\
	    Windows Workstation & 2 & Internet Explorer, JDK, Office,\\
	    & & DirectX, Edge\\
	    Windows Workstation & 2 & LiveMeeting, Edge\\
	    Windows Workstation & 1 & Internet Explorer, Office,\\ 
	    & & Chrome, ExpressionWeb, JScript \\
	    Windows Server 2008 & 1 & -\\
	    SMB Device & 2 & -\\
	    Ubuntu Machine & 2 & Pidgin, Chrome, Firefox, Apport,\\
	    & & Python, Jasper, OpenSSL,\\
	    Red Hat Enterprise  & 1 & Enterprise MRG, Evince\\
	    Machine & & Libxml2, Poppler\\
	    Linux Database Server & 1 & -\\
	    TWiki Web Server & 1 & TWiki, PCRE, PHP, Samba\\
	    Remote Login Machine & 1 & OpenSSH\\
        \bottomrule
    \end{tabular}
\end{table}
\end{document}